\documentclass{jpp}

\usepackage{graphicx,bm,color}
\usepackage{natbib}
\usepackage{color}

\graphicspath{{./fig/}{./png/}}

\newcommand{\bec}[1]{\mbox{\boldmath $ #1$}}
\newcommand{\EQ}{\begin{equation}}
\newcommand{\EN}{\end{equation}}
\newcommand{\EQA}{\begin{eqnarray}}
\newcommand{\ENA}{\end{eqnarray}}

\newcommand{\Fig}[1]{Fig.~\ref{#1}}

\newcommand{\meanrho}{\overline{\rho}}

{}
{}
{}

{}
{}
\newcommand{\meanEMF}{\overline{\mbox{\boldmath ${\cal E}$}}{}}{}
{}
{}
{}
{}
{}
\newcommand{\meanBB}{\overline{\mbox{\boldmath $B$}}{}}{}
{}
{}
{}
{}
{}
{}
{}
{}
\newcommand{\meanUU}{\overline{\bm{U}}}

{}
{}
{}

\newcommand{\meanB}{\overline{B}}

\newcommand{\meanp}{\overline{p}}

\newcommand{\meanT}{\overline{T}}

{}

{}
{}

\newcommand{\uu}{\mbox{\boldmath $u$} {}}
\newcommand{\UU}{\mbox{\boldmath $U$} {}}

\newcommand{\BB}{\mbox{\boldmath $B$} {}}

\newcommand{\nab}{\mbox{\boldmath $\nabla$} {}}

\newcommand{\SSSS}{\mbox{\boldmath ${\sf S}$} {}}

\newcommand{\dive}{{\rm div}  \, {}}

\def\etat{\eta_{\rm t}}
\def\Dt{D_{\rm t}}

\def\half{{\textstyle{1\over2}}}

\def\onethird{{\textstyle{1\over3}}}

\title[Compressibility in turbulent MHD and passive scalar transport]
{Compressibility in turbulent MHD and passive scalar transport: mean-field theory}

\author[I. Rogachevskii, N. Kleeorin and A. Brandenburg]%
{I.\ns R\ls O\ls G\ls A\ls C\ls H\ls E\ls V\ls S\ls K\ls I\ls I$^{1,2}$
  \thanks{Email address for correspondence: gary@bgu.ac.il},
N.\ns K\ls L\ls E\ls E\ls O\ls R\ls I\ls N$^{1,2}$
\and
A.\ns B\ls R\ls A\ls N\ls D\ls E\ls N\ls B\ls U\ls R\ls G$^{2,3,4}$
}

\affiliation{
$^{1}$Department of Mechanical Engineering, Ben-Gurion University of
the Negev, P. O. Box 653, 84105 Beer-Sheva, Israel\\
$^{2}$Nordita, KTH Royal Institute of Technology
and Stockholm University, Roslagstullsbacken 23,
10691 Stockholm, Sweden\\
$^{3}$Laboratory for Atmospheric and Space Physics,
JILA and Department of Astrophysical and Planetary Sciences,
University of Colorado, Boulder, CO 80303, USA\\
$^{4}$Department of Astronomy, AlbaNova University Center, Stockholm
University, SE-10691 Stockholm, Sweden
}

\date{\today,~ $ $Revision: 1.118 $ \!\! $}
\begin{document}

\maketitle

\begin{abstract}
We develop a mean-field theory of compressibility effects in
turbulent magnetohydrodynamics and passive scalar transport
using the quasi-linear approximation and the spectral $\tau$-approach.
We find that compressibility decreases
the $\alpha$ effect and the turbulent magnetic diffusivity
both at small and large magnetic Reynolds numbers, Rm.
Similarly, compressibility decreases the
turbulent diffusivity for passive scalars both at small and large
P\'eclet numbers, Pe.
On the other hand, compressibility does not affect the effective
pumping velocity of the magnetic field for large Rm, but it decreases
it for small Rm.
Density stratification causes turbulent pumping of passive scalars,
but it is found to become weaker with increasing compressibility.
No such pumping effect exists for magnetic fields.
However, compressibility results in a new passive scalar pumping effect
from regions of low to high turbulent intensity
both for small and large P\'eclet numbers.
It can be interpreted as compressible turbophoresis
of noninertial particles and gaseous admixtures, while the classical
turbophoresis effect exists only for inertial particles and causes
them to be pumped to regions with lower turbulent intensity.
\end{abstract}

\section{Introduction}

The generation of magnetic fields by a turbulent flow of a conducting
fluid is a fundamental problem which has a number of applications
in astrophysics, geophysics, planetary
physics, and laboratory experiments \citep[see, e.g.,][]
{moffatt1978,parker1979,krause1980,zeldovich1983,ruzmaikin1988,brandenburg2005,ruediger2013}.
Two types of turbulent dynamos are usually considered:
large-scale and small-scale dynamos.
Magnetic field generation on scales much smaller and much larger than the
integral scale of turbulence are described as
small-scale dynamo and large-scale dynamo, respectively.

The theory of large-scale dynamos
is based on a mean-field approach \citep{moffatt1978,krause1980}
in which the magnetic and velocity fields are decomposed into mean
and fluctuating parts.
Assuming that the averages obey the Reynolds rules,
it follows that the fluctuating parts have zero mean values.
Let us furthermore assume that there exists a
separation of scales, i.e., the maximum scale of
turbulent motions (the turbulent integral scale) is much smaller than
the characteristic scales of inhomogeneity of the mean fields.
In the framework of the mean-field approach, turbulence effects
are described in terms of the mean electromotive force,
$\meanEMF=\overline{{\bm u} \times {\bm b}}$ as a function of the large-scale
magnetic field $\meanBB$, where ${\bm u}$ and ${\bm b}$ are fluctuations
of velocity and magnetic field.

For a number of astrophysical applications, the nonlinear dependence
of the function $\meanEMF(\meanBB)$ has been determined analytically
for flows that are incompressible, i.e., div
${\bm \UU}=0$ \citep[see, e.g.,][]{
rogachevskii2000,rogachevskii2001,rogachevskii2004,
rogachevskii2012,ruediger2013,rogachevskii2017},
or for low-Mach-number density-stratified flows in the anelastic approximation,
$\dive(\rho \, \UU)=0$ \citep[see, e.g.,][]{kleeorin2003,rogachevskii2006},
where $\rho$ is the fluid density.
However, in many astrophysical flows, the Mach number is not small
and compressibility effects on the mean electromotive force can be important.

Similar questions associated with compressibility effects arise in turbulent
transport of passive scalars, where the mean-field approach
allows us to determine the large-scale dynamics of passive scalars
in small-scale turbulence \citep{mccomb1990,zeldovich1990,frisch1995,Zaichik2008,monin2013}.
The effects of turbulence on passive scalar transport are described
by means of a turbulent flux of passive scalar concentration or particle number density
${\bm F}=\overline{n \, {\bm u}}$, where $n$ is the fluctuation of the particle
number density \citep{csanady1980,crowe2011,balachandar2010,piterbarg2013}.
Compressibility effects play a crucial role in particle transport,
for example they can cause the formation of spatially inhomogeneous
distributions of particles, known as particle clusters.
These effects are very important in a range of astrophysical
and planetary science applications,
where turbulence can be temperature stratified at finite Mach numbers
\citep{armitage2010,ruzmaikin1988,priest1982}.

For temperature-stratified turbulence, large-scale particle clusters
(on scales much larger than the integral scale of the turbulence)
are formed due to turbulent thermal diffusion
\citep{elperin1996,elperin1997}.
The effects related to compressibility of the turbulent flow or the particle
velocity field, result in a pumping effect of particles in regions
of minimum mean fluid temperature.
In particular, turbulent thermal diffusion causes a turbulent non-diffusive
flux of particles in the direction of the turbulent heat flux, so that
particles are accumulated in the vicinity of the mean temperature minimum.
The phenomenon of turbulent thermal diffusion has
been studied theoretically \citep{elperin2000,elperin2001,amir2017},
detected in direct numerical simulations \citep{haugen2012},
different laboratory experiments \citep{buchholz2004,eidelman2006,eidelman2010,amir2017},
and atmospheric turbulence with temperature inversions \citep{sofiev2009}.
It was also shown to be important for concentrating dust in protoplanetary
discs \citep{hubbard2015}.

A suppression of turbulent magnetic diffusivity by the compressibility of
a random homogeneous flow using
the quasi-linear approach (the second-order correlation
approximation) was demonstrated by \cite{krause1980}.
In particular, \cite{krause1980} derived an equation for
the turbulent magnetic diffusivity, $\etat$, for small magnetic Reynolds numbers:
$\etat=(\overline{\bec{\psi}^2} - \overline{\phi^2})/3 \eta$,
where $\eta$ is the microscopic magnetic diffusivity
and the velocity fluctuations are represented as the sum of vortical
and potential parts, ${\bm u}=\bec{\nabla} {\bm \times} \bec{\psi}
+ \bec{\nabla} \phi$.
Later, \cite{radler2011} determined mean-field diffusivities both
for passive scalars and magnetic fields
for an irrotational homogeneous deterministic flow,
using the quasi-linear approach
and the test-field method in direct numerical simulations.
They showed that the expression for the turbulent diffusivity
of a passive scalar coincides with that of $\etat$
for small magnetic Reynolds and P\'eclet numbers
after replacing $\eta$ by the molecular diffusion coefficient
for the passive scalar.
They found that the total mean-field
diffusivity in irrotational flows may well be smaller than the molecular
diffusivity.

In the present study, a mean-field theory of compressibility effects in
turbulent magnetohydrodynamics (MHD) and passive scalar transport
is developed for inhomogeneous density-stratified
turbulent flows at arbitrary Mach number.
We also consider helical turbulence with uniform
kinetic helicity.
We use the quasi-linear approach,
which is valid for small magnetic Reynolds and P\'eclet numbers,
and the spectral tau approach, which is applicable to large magnetic Reynolds
and P\'eclet numbers in fully developed turbulence.
This allows us to determine the mean electromotive force
and the turbulent flux of particles in compressible density-stratified
inhomogeneous turbulence.

\section{Turbulent transport of magnetic field}

In this section we study turbulent transport of a
magnetic field in compressible helical inhomogeneous
turbulence for small and large magnetic Reynolds
numbers.

\subsection{Governing equations}

The magnetic field $\BB({\bm x},t)$ is governed by
the induction equation:
\begin{eqnarray}
{\partial \BB \over \partial t} = \bec{\nabla} \times
(\UU {\bm \times} \BB - \eta \bec{\nabla} {\bm \times} \BB) ,
\label{TC0}
\end{eqnarray}
where $\eta$ is the magnetic
diffusivity due to the electrical conductivity of the fluid,
and fluid velocity $\UU({\bm x},t)$ is determined by the Navier-Stokes equation,
\begin{eqnarray}
\rho \, \left({\partial \over
\partial t} + \UU \cdot \nab\right)\UU - \nab\cdot\left(2\nu \rho \SSSS^{\rm(U)}\right) =
{1 \over\mu_0} (\bec{\nabla} \times \BB) \times \BB
-\bec{\nabla} P,
\label{CCC1}
\end{eqnarray}
where $P({\bm x},t)$ is the fluid pressure,
${\sf S}_{ij}^{\rm(U)} = \half(U_{i,j}+U_{j,i}) - \onethird\delta_{ij}\nab\cdot\UU$
are the components of the traceless rate-of-strain-tensor $\SSSS^{\rm(U)}$,
commas denote partial differentiation,
and $\nu$ is the kinematic viscosity.

We apply a mean-field approach and use Reynolds
averaging \citep{krause1980}.
In the framework of this approach, velocity and magnetic fields, fluid density,
and pressure are decomposed into mean (denoted by overbars) and fluctuating parts (lowercase symbols).
Ensemble averaging of (\ref{TC0}) yields
an equation for the mean magnetic field $\meanBB({\bm x},t)$:
\begin{eqnarray}
{\partial \meanBB \over \partial t} = \bec{\nabla} \times
(\meanUU {\bm \times} \meanBB + \meanEMF
- \eta \bec{\nabla} {\bm \times} \meanBB) ,
\label{C1}
\end{eqnarray}
where $\meanUU({\bm x},t)$ is the mean velocity
and $\meanEMF=\overline{{\bm u} \times {\bm b}}$
is the mean electromotive force.
For simplicity, we consider in this study the case
$\meanUU={\bm 0}$.
We determine $\meanEMF$ for compressible inhomogeneous and helical turbulence.
The procedure of the derivation of the equation for the mean
electromotive force is as follows.
The momentum and induction equations for fluctuations are given by
\begin{eqnarray}
&& \overline{\rho} \, {\partial {\bm u} \over
\partial t} - \nab\cdot\left(2\nu\overline{\rho}\SSSS^{\rm(u)}\right) =
{1 \over\mu_0} [({\bm b} {\bm \cdot}
\bec{\nabla}) \overline{\bm B} + (\overline{\bm B}
{\bm \cdot} \bec{\nabla}){\bm
b}] -\bec{\nabla} p_{\rm tot} + {\bm u}^{\rm(N)} ,
\label{D1} \\
&& {\partial {\bm b} \over \partial t}
- \eta \Delta {\bm b} =(\overline{\bm B}
{\bm \cdot} \bec{\nabla}){\bm u} - ({\bm u} {\bm \cdot}
\bec{\nabla}) \overline{\bm B}
- \overline{\bm B}
\, (\bec{\nabla} {\bm \cdot} {\bm u}) + {\bm b}^{\rm(N)} ,
\label{D2}
\end{eqnarray}
where ${\bm u}({\bm x},t)$ and ${\bm b}({\bm x},t)$ are the fluctuations of
velocity and magnetic fields,
${\bm u}^{\rm(N)}$ and ${\bm b}^{\rm(N)}$
are terms that are nonlinear in the fluctuations,
$p_{\rm tot} = p + \mu_0^{-1} \,(\overline{\bm B}
{\bm \cdot} {\bm b}) $ are the fluctuations of the total pressure,
$\SSSS^{\rm(u)} \equiv {\sf S}_{ij}^{\rm(u)} = \half(u_{i,j}+u_{j,i}) -
\onethird\delta_{ij}\nab\cdot{\bm u}$, and $p({\bm x},t)$ is the fluctuation
of the fluid pressure.
We rewrite (\ref{D1}) and (\ref{D2}) in ${\bm k}$ space.
In the following, we derive equations for
the mean electromotive force for small and large
magnetic Reynolds numbers.

\subsection{Small magnetic Reynolds numbers}
\label{small-Rm}

We use a quasi-linear approach
(also known in the literature as ``first-order
smoothing'' or ``second-order correlation approximation''),
which is valid for small hydrodynamic and magnetic Reynolds numbers,
${\rm Re}=\ell_0 \, \sqrt{\overline{{\bm u}^2}} /\nu \ll 1$ and
${\rm Rm}=\ell_0 \, \sqrt{\overline{{\bm u}^2}} /\eta \ll 1$, respectively.
Here $\sqrt{\overline{{\bm u}^2}}$ is the characteristic turbulent
velocity in the integral turbulent scale $\ell_0$.
In the high conductivity limit
(very small microscopic magnetic diffusivity $\eta$),
the quasi-linear approach is only
valid for small Strouhal numbers,
which is defined as the ratio of turbulent correlation time,
$\tau_0$, to turnover time, $\ell_{0} / \sqrt{\overline{{\bm u}^2}}$.
This implies that the quasi-linear approach in this limit is only
valid for very short turbulent correlation times.
Often \citep[see, e.g.,][]{krause1980}
the quasi-linear approach is applied only to
the induction equation and the origin of the velocity field
is not discussed. In this case the evolution of the magnetic field
is considered for a prescribed velocity field, and formally
the validity of the quasi-linear approach only requires small
magnetic Reynolds numbers.

\subsubsection{Multi-scale approach}
\label{Multi-scale}

To determine the mean electromotive force, which is a one-point correlation function,
and to take into account small-scale properties of the turbulence,
e.g., the turbulent spectrum, one must use two-point correlation
functions. For fully developed turbulence, scalings for
the turbulent correlation time and energy spectrum are related via
the Kolmogorov scalings \citep{mccomb1990,frisch1995,monin2013}. This is the reason why
for calculations of the mean electromotive force in fully developed turbulence,
we use instantaneous two-point correlation
functions. On the other hand, for a random flow with small
${\rm Re}$ and ${\rm Rm}$, there are no universal scalings for
the correlation time and energy spectrum.
This is the reason why we have to use
non-instantaneous two-point correlation
functions in this case.

In the framework of the mean-field approach, we assume
that there is a separation of spatial and temporal scales, i.e.,
$\ell_0 \ll L_B$ and $\tau_0 \ll t_B$, where
$L_B$ and $t_B$ are the characteristic spatial and temporal scales
characterizing the variations of the mean magnetic field.
The mean fields depend on ``slow'' variables, while fluctuations
depend on ``fast'' variables.
Separation into slow and fast variables is widely used in theoretical physics,
and all calculations are reduced to Taylor expansions of all functions
using small parameters $\ell_0 / L_B$ and $\tau_0 / t_B$.
The findings are further truncated to leading order terms.

Separation to slow and fast variables
is performed by means of a standard multi-scale approach
\citep{roberts1975}.
In the framework of this approach,
the non-instantaneous two-point second-order correlation
function of ${\bm b}$ and ${\bm u}$ is written as follows:
\begin{eqnarray}
\overline{b_i({\bm x},t_1) \, u_j ({\bm  y},t_2)}
&=& \int \,d\omega_1 \, d\omega_2 \,d{\bm k}_1 \, d{\bm k}_2 \,\,
\overline{b_i({\bm k}_1,\omega_1) u_j({\bm k}_2,\omega_2)} \, \exp
\big[i({\bm  k}_1 {\bm \cdot} {\bm x}
+{\bm k}_2 {\bm \cdot}{\bm y})
\nonumber\\
&& + i(\omega_1 t_1 + \omega_2 t_2)\big]
= \int g_{ij}({\bm k},\omega,t,{\bm R})  \exp[i {\bm k}
{\bm \cdot} {\bm r} + i\omega \, \tilde \tau] \,d\omega \,d
{\bm k} ,
\label{E2}
\end{eqnarray}
where
\begin{eqnarray}
g_{ij}({\bm k},\omega,{\bm R},t) &=&
 \int \overline{b_i({\bm k}_1,\omega_1) \, u_j({\bm k}_2,\omega_2)}
\exp[i \Omega t+ i {\bm K} {\bm \cdot} {\bm R}] \,d \Omega \,d {\bm  K}.
\label{E3}
\end{eqnarray}
Here we introduced large-scale variables: ${\bm R} = ({\bm x}
+ {\bm y}) / 2$, $\, {\bm K} = {\bm k}_1 + {\bm k}_2$,
$\, t = (t_1 + t_2) / 2$, $\, \Omega = \omega_1 + \omega_2$,
and small-scale variables: ${\bm r} = {\bm x} - {\bm y}$,
$\, {\bm k} = ({\bm k}_1 - {\bm k}_2) / 2$, $\, \tilde \tau
= t_1 - t_2$, $\, \omega = (\omega_1 - \omega_2) / 2$.
This implies that $\omega_1 = \omega + \Omega / 2$, $\, \omega_2 =
- \omega + \Omega / 2$, ${\bm k}_1 = {\bm k} + {\bm  K} / 2$,
and ${\bm k}_2 = - {\bm k} + {\bm  K} / 2$.
Mean-fields depend on the large-scale variables, while
fluctuations depend on the small-scale variables.
We have used here the Fourier transformation:
\begin{eqnarray}
\Phi({\bm x},t) = \int \Phi({\bm k}, \omega) \exp\big[i({\bm  k}
{\bm \cdot} {\bm x} + \omega t)\big] \,d\omega \,d{\bm k} .
\label{EEE3}
\end{eqnarray}
Similarly to (\ref{E2})-(\ref{E3}), we obtain
\begin{eqnarray}
f_{ij}({\bm k},\omega,{\bm R},t) &=&
 \int \overline{u_i({\bm k}_1,\omega_1) \, u_j({\bm k}_2,\omega_2)}
\exp[i \Omega t+ i {\bm K} {\bm \cdot} {\bm R}] \,d \Omega \,d {\bm  K} .
\label{EEEE3}
\end{eqnarray}
After separation into slow and fast variables and calculating the function
$g_{ij}({\bm k},\omega,{\bm R},t)$, expression~(\ref{E2}) allows us to
determine the cross-helicity tensor in physical space and to calculate the
limit of ${\bm r} \to {\bm 0}$ and $\tilde \tau \to 0$, which yields
\begin{eqnarray}
\overline{b_i({\bm x},t) \, u_j ({\bm  x},t)} = \int g_{ij}({\bm k},\omega,{\bm R},t)
\,d\omega \,d {\bm k} .
\label{EEE3}
\end{eqnarray}
For brevity of notations we omit below the large-scale variables $t$ and ${\bm R}$
in the correlation functions $f_{ij}({\bm k},\omega,{\bm R},t)$ and
$g_{ij}({\bm k},\omega,{\bm R},t)$.

\subsubsection{Cross-helicity tensor}

In the framework of the quasi-linear approach,
we neglect nonlinear terms, but keep molecular dissipative
terms in (\ref{D2}) for the magnetic fluctuations.
This allows us to obtain the solution of (\ref{D2}) in the
limit of small magnetic Reynolds number.
Using this solution, we derive the equation for the correlation function
$g_{ij}({\bm k},\omega)$:
\begin{eqnarray}
&& g_{ij}({\bm k},\omega) = G_\eta \, \biggl\{\meanB_p \biggl[i k_p f_{ij}
+ {1 \over 2} \nabla_p f_{ij} - \eta k^2 G_\eta \, k_{sp} \nabla_s f_{ij}\biggr]
- \meanB_i \biggl[i k_p f_{pj}
\nonumber\\
&& \quad \quad + {1 \over 2} \nabla_p f_{pj} - \eta k^2 G_\eta \, k_{sp} \nabla_s f_{pj}\biggr] - (\nabla_s \meanB_p) \biggl[ {1 \over 2} k_p {\partial f_{ij} \over \partial k_s}
+ \eta k^2 G_\eta \, k_{sp} f_{ij}\biggr]
\nonumber\\
&& \quad \quad + {1 \over 2} (\nabla_s \meanB_i) \biggl[- f_{sj} +  k_p {\partial f_{pj} \over \partial k_s} + 2\eta k^2 G_\eta \, k_{sp} f_{pj}\biggr] \biggl\},
\label{G2}
\end{eqnarray}
where $k_{ij}=k_i \, k_j / k^2$ and $G_\eta \equiv G_\eta({\bm k},\omega)
= (\eta {\bm k}^2 + i \omega)^{-1}$.
The derivation of (\ref{G2}) is given in Appendix~\ref{appA}.
Here, for brevity of notations, we omit the large-scale variables ${\bm R}$ and $t$
in the mean magnetic field $\meanB_p$ and correlation functions $g_{ij}$ and $f_{ij}$.
When $\meanBB^2/ 4 \pi \ll \meanrho \overline{{\bm u}^2}$, the correlation function
$f_{ij}({\bm k},\omega)$ in (\ref{G2}) should be replaced by
the correlation function $f_{ij}^{(0)}({\bm k},\omega)$
given below by (\ref{B5}) for the random flow
with a zero mean magnetic field, called background flow.

\subsubsection{A model for the background random flow for Re $ \ll 1$}

In the next step of the derivation, we need
a model for the background random flow with a zero mean magnetic field.
We consider a random, statistically stationary,
density-stratified, inhomogeneous, compressible
and helical background flow,
which is determined by the following correlation function
in Fourier space:
\begin{eqnarray}
f_{ij}^{(0)}({\bm k},\omega,{\bm R}) &=&
{\Phi_u(\omega) \, E(k) \over 8 \pi \, k^2 \,
(1+ \sigma_c)} \biggl\{\Big[\delta_{ij} - k_{ij} + {i \over k^2} \,
\big(k_j \lambda_i - k_i \lambda_j\big) + 2 \sigma_c \, k_{ij}
\nonumber\\
&& + (1 +2 \sigma_c) \, {i \over 2 k^2} \,
\big(k_i \nabla_j - k_j \nabla_i\big)  \Big] \, \overline{{\bm u}^2}
- {i \over k^2} \, \varepsilon_{ijp} \, k_p \chi \biggr\} .
\label{B5}
\end{eqnarray}
Here ${\bm \lambda} = - {\bm \nabla} \ln \meanrho$ characterizes the
fluid density stratification, $\delta_{ij}$ is the Kronecker tensor,
$\varepsilon_{ijn}$ is the fully antisymmetric
Levi-Civita tensor, $\chi=\overline{{\bm u} \,{\bf\cdot}  \,(\bec{\nabla}
{\bf \times} \, {\bm u})}$ is the kinetic helicity, and the parameter
\begin{eqnarray}
\sigma_c = {\overline{ (\bec{\nabla} \cdot \, {\bm u})^2 }
\over \overline{(\bec{\nabla} \times {\bm u})^{2} }}
  \label{B40}
\end{eqnarray}
is the degree of compressibility of the turbulent
velocity field.
In (\ref{B5}),  we neglect higher order effects
$\sim$O$(\lambda^2 \, \overline{{\bm u}^2}, \nabla^2 \, \overline{{\bm u}^2},
\lambda_i \nabla_i\overline{{\bm u}^2})$,
i.e., we assume that $\ell_0 \ll H$ and $\ell_0 \ll L_B$,
where $\ell_{0}$ is the maximum scale of random motions,
$H=|{\bm \lambda}|^{-1}$ is the mean density variation scale,
which is assumed to be constant.
This means that in (\ref{B5}) we only take into account leading effects, which are linear
in stratification ($\propto \ell_0 / H$) and inhomogeneity of turbulence
($\propto \ell_0 / L_B$).
Generally, stratification also contributes to div~${\bm u}$, and
therefore it contributes to the parameter $\sigma_c$.
However, this contribution is small [$\sim$O$(\lambda^2 \, \overline{{\bm u}^2})$],
and neglected in (\ref{B5}).
This implies that we separate effects of the arbitrary Mach number, characterized by
the parameter $\sigma_c$, and density stratification, described by ${\bm \lambda}$.
Note that the degree of compressibility, $\sigma_c$,
depends on the Mach number, but this
dependence is not known for arbitrary Mach numbers and has
to be determined in direct numerical simulations.

We also do not consider here effects related to the non-uniformity of kinetic helicity
($\propto {\bm \nabla}\chi$)
or the combined effect of stratification and kinetic helicity
($\propto {\bm \lambda}\chi$), because
their contributions to the mean electromotive force are
much smaller in comparison with the standard contributions to
the mean electromotive force caused by the turbulent motions
(they are of the order of the terms we neglected in the present study).
The non-uniformity of kinetic helicity \citep{yokoi1993,yokoi2016,kleeorin2018}
or the combined effect of stratification and kinetic helicity \citep{kleeorin2018}
contribute to the generation of large-scale vorticity
or large-scale shear motions.

In (\ref{B5}), we assume that helical and non-helical parts of random flow have the same
power-law spectrum $E(k) = k_0^{-1} \, (q-1) \, (k / k_{0})^{-q}$ for the
wave number range $k_0<k<k_\nu$, where $k_{\nu} = 1 / \ell_{\nu}$
is the wave number based on the viscous scale $\ell_{\nu}$,
and $k_{0} = 1 / \ell_{0} \ll k_\nu$.
We assume that there are no random motions for
$k < k_0$. This implies that $\ell_0$ is the maximum scale
of random motions, and $E(k)=0$ for $k < k_0$.
For simplicity we also assume that
compressible and incompressible parts of the
random flow have the same spectra.
We consider the frequency function $\Phi_u(\omega)$ in the form of a
Lorentz profile: $\Phi_u(\omega)=[\pi \tau_0 \,
(\omega^2 + \tau_0^{-2})]^{-1}$,
where $\tau_0 = \ell_{0} / \sqrt{\overline{ {\bm u}^2 }}$ is the correlation time and
$\sqrt{\overline{ {\bm u}^2 }}$ is the characteristic turbulent
velocity at scale $\ell_{0}$. This model for
the frequency function corresponds to the
correlation function
\begin{eqnarray}
\overline{u_i(t) u_j(t+\tau)} \propto \exp (-\tau / \tau_0).
\end{eqnarray}
We take into account that for small magnetic Reynolds numbers,
$\tau_0 \gg (\eta k^2)^{-1}$ for all turbulent scales.

In spite of the large-scale effects caused by stratification ${\bm \lambda}$ and inhomogeneous turbulence ${\bm \nabla}\overline{{\bm u}^2}$, the model~(\ref{B5}) describes a weakly anisotropic approximation. Indeed, the contributions of these effects to $f_{ij}^{(0)}$ are small because $\ell_0 \ll H$ and $\ell_0 \ll L_B$.
Since we consider only linear effects in ${\bm \lambda}$ and ${\bm \nabla}\overline{{\bm u}^2}$, the second rank tensor $f_{ij}^{(0)}$ is constructed as a linear combination of symmetric tensors with respect to the indexes $i$ and $j$, $\delta_{ij}$, $k_{ij}$, non-symmetric tensors: $k_i \lambda_j$, $ k_j \lambda_i$, $k_i \nabla_j\overline{{\bm u}^2}$, $ k_j \nabla_i\overline{{\bm u}^2}$, and fully antisymmetric tensor $\varepsilon_{ijp} \, k_p$.
To determine unknown coefficients multiplying by these tensors,
we use the following conditions in the derivation of (\ref{B5}):
\\
(i) $\int f_{ii}^{(0)} ({\bm k},\omega,{\bm K}) \, \exp(i {\bm K} {\bm \cdot} {\bm R}) \, d {\bm k} \, d\omega \, d {\bm K} = \overline{{\bm u}^2}$;
\\
(ii) $i \varepsilon_{ipj} \int (-k_p + K_p/2) \, f_{ij}^{(0)} ({\bm k},\omega,{\bm K}) \, \exp(i {\bm K} {\bm \cdot} {\bm R}) \, d {\bm k} \, d\omega \, d {\bm K} = \chi$;
\\
(iii) $f_{ij}^{(0)} ({\bm k},\omega,{\bm K}) = f_{ji}^{*(0)} ({\bm k},\omega,{\bm K}) = f_{ji}^{(0)} (-{\bm k},\omega,{\bm K})$.
\\
(iv) $\int (k_i + K_i/2) \, (k_j - K_j/2) \, f_{ij}^{(0)} ({\bm k},\omega,{\bm K}) \, \exp(i {\bm K} {\bm \cdot} {\bm R}) \, d {\bm k} \, d\omega \, d {\bm K} = \overline{\left({\rm div} \, {\bm u}\right)^2}$;
\\
(v) for very low Mach number and when the parameter $\sigma_c$ is very small,
the continuity equation can be written in the anelastic approximation,
${\rm div} \, (\meanrho \, {\bm u}) = 0$,
which implies that
$(ik_i + iK_i/2 - \lambda_i) f_{ij}^{(0)}({\bm k},\omega,{\bm K}) = 0$ and
$(-ik_j + iK_j/2 - \lambda_j) f_{ij}^{(0)} ({\bm k},\omega,{\bm K})= 0$.

Different contributions to (\ref{B5}) have been discussed by
\cite{batchelor1953}, \cite{elperin1995}, \cite{radler2003},
\cite{rogachevskii2018}, and \cite{kleeorin2018}.
Note that for a purely potential flow, Eq. (2.12) for the background turbulence
is used in the limit $\sigma_c \to \infty$, which yields
\begin{eqnarray}
f_{ij}^{(0)}({\bm k},\omega,{\bm R}) =
{\Phi_u(\omega) \, E(k) \over 4 \pi \, k^2}
\biggl[k_{ij} + {i \over 2 k^2} \,
\big(k_i \nabla_j - k_j \nabla_i\big)  \biggr] \, \overline{{\bm u}^2} .
\label{B55}
\end{eqnarray}

\subsubsection{Mean electromotive force for ${\rm Rm} \ll 1$}

The mean electromotive force is given by
\begin{eqnarray}
\overline{\cal E}_{i} &=& \varepsilon_{inm} \, \int
\overline{b_m({\bm k},\omega) \,
u_n(-{\bm k},-\omega)}\, \,d {\bm k} \,d \omega = \varepsilon_{inm} \, \int
g_{mn}({\bm k},\omega) \,d {\bm k} \,d \omega .
 \label{G3}
\end{eqnarray}
Integrating in $\omega$ space and ${\bm k}$ space
in (\ref{G3})
and neglecting higher order effects
$\sim$O$(\lambda^2 \, \overline{{\bm u}^2}$, $\nabla^2 \, \overline{{\bm u}^2},
\lambda_i \nabla_i\overline{{\bm u}^2})$,
we arrive at an equation
for the mean electromotive force, $\meanEMF
=\overline{{\bm u} \times {\bm b}}$,
for compressible, density-stratified, inhomogeneous and helical background turbulence:
\begin{eqnarray}
\meanEMF = \alpha \meanBB + \bm{\gamma}
\times \meanBB - \etat \, \bm{\nabla} {\bm \times}
\meanBB ,
\label{C2}
\end{eqnarray}
where the $\alpha$ effect, the
turbulent magnetic diffusivity $\etat$, and
the pumping velocity $\bm{\gamma}$ for ${\rm Rm} \ll 1$ are given by
\begin{eqnarray}
\alpha &=& - {(q-1)\over 3(q+1)} \, {\tau_0 \, \overline{{\bm u} \,{\bf\cdot}  \,(\bec{\nabla} {\bf \times} \, {\bm u})} \over 1 + \sigma_c}  \,  {\rm Rm},
\label{C0}\\
\etat &=& {(q-1)\over 3(q+1)} \, \tau_0 \, \overline{{\bm u}^2} \, \left(1 - {2 \sigma_c
\over 1 + \sigma_c} \right) \,{\rm Rm}\,
\equiv {(q-1)\over 3(q+1)} \, \left({1- \sigma_c
\over 1 + \sigma_c} \right) \,{\rm Rm}^2 \, \eta,
\label{C3}\\
\bm{\gamma} &=& - {1 \over 2} \, {\bm \nabla} \etat ,
\label{C4}
\end{eqnarray}
and $\tau_0 = \ell_0 /\sqrt{\overline{{\bm u}^2}}$.
Let us analyze the obtained results.
Since $\tau_0 \, {\rm Rm} = \ell_0^2 /\eta$, the turbulent transport coefficients
given by (\ref{C0})--(\ref{C4}) are determined only by the resistive time scale,
$\ell_0^2 /\eta$.
The effective pumping velocity $\bm{\gamma}$ of the large-scale magnetic field
is independent of ${\bm \lambda}$
(see detailed derivation of (\ref{C4}) and discussion in Appendix~\ref{appC}).
We also have shown that there are no contributions of $\bm{\lambda}$ to the effective
pumping velocity even for strong stratifications (see Appendix~\ref{appC}).

Equations~(\ref{C0})--(\ref{C4}) imply that for small
magnetic Reynolds numbers, compressibility effects
characterized by the parameter $\sigma_c$
decrease the $\alpha$ effect, turbulent magnetic diffusion, and
the effective pumping velocity.
However, the total magnetic diffusivity, $\eta + \etat$, cannot be negative,
because for ${\rm Rm} \ll 1$ the magnetic diffusivity, $\eta$,
is much larger than the turbulent value, i.e., $\eta \gg |\etat|$.
The decrease of turbulent magnetic diffusivity
by compressible flows is consistent with the results obtained by
\cite{krause1980} and \cite{radler2011} for homogeneous random flow
at small magnetic Reynolds numbers.
Detailed comparison with these results is discussed in section~\ref{comparison}.

As follows from (\ref{C2})--(\ref{C3}), the resulting mean electromotive force
$\overline{\cal E}_{i}$ is determined by the isotropic $\alpha$ effect
and turbulent diffusivity.
The reason for that is as follows.
(i) We consider the case of weak mean magnetic field,
i.e., the energy of the mean magnetic field is much smaller
than the turbulent kinetic energy.
For large mean magnetic field, when the energy of the mean magnetic field
is of the order of the turbulent kinetic energy,
the $\alpha$ tensor and turbulent diffusivity
are anisotropic \citep{rogachevskii2000,rogachevskii2001}.
(ii) We do not consider the effects of uniform rotation or large-scale shear.
In the case of rotating turbulence \citep{radler2003,kleeorin2003}
or turbulence with large-scale shear \citep{rogachevskii2003,rogachevskii2004},
the $\alpha$ tensor and turbulent diffusivity are anisotropic.
(iii) For anisotropic background turbulence, e.g., for turbulent
convection \citep{kleeorin2003}, or for magnetically driven turbulence
with relativistic particles \citep{rogachevskii2012,rogachevskii2017},
the $\alpha$ tensor and turbulent diffusivity are also anisotropic.
(iv) The model (\ref{B5}) of background random flow is weakly anisotropic,
because $\ell_0 \ll H$ and $\ell_0 \ll L_B$.

\subsection{Large fluid and magnetic Reynolds numbers}
\label{large-Rm}

In this section we consider the case of
large fluid and magnetic Reynolds numbers, so that
turbulence is fully developed, the Strouhal number
is of the order of unity,
and the turbulent correlation time is scale-dependent, like
in Kolmogorov type turbulence
\citep[see, e.g.,][]{monin2013,mccomb1990,frisch1995}.
In this case, we perform the Fourier transformation only in
${\bm k}$ space (but not in $\omega$ space),
as is usually done in studies of turbulent transport in a fully developed Kolmogorov-type turbulence.
We take into account the nonlinear terms in (\ref{D1})
and~(\ref{D2}) for velocity and magnetic fluctuations
and apply the $\tau$ approach.

The $\tau$ approach is a universal tool in turbulent transport
for strongly nonlinear systems that allow us to obtain closed results
and compare them with the results of laboratory experiments, observations,
and numerical simulations
\citep[see, e.g.,][]{orszag1970,pouquet1976,kleeorin1990,rogachevskii2004,brandenburg2005,rogachevskii2007,rogachevskii2011}.
The $\tau$ approximation reproduces many
well-known phenomena found by other methods in turbulent transport
of particles \citep[see, e.g.,][]{elperin1996,elperin1997,pandya2002,blackman2003,reeks2005,sofiev2009}
and magnetic fields, in turbulent convection \citep[see, e.g.,][]{elperin2002,elperin2006} and stably stratified turbulent flows \citep[see, e.g.,][]{elperin2002,zilitinkevich2009,zilitinkevich2013} for large fluid and magnetic Reynolds and P\'{e}clet numbers.
This approach is different from the quasi-linear approach
applied in the high-conductivity limit. The latter approach
is only valid for small Strouhal numbers, i.e., for very
short correlation time. Therefore, the final results
found with this approach are different from those
obtained using the $\tau$ approach.

Using (\ref{D2}) for the magnetic fluctuations ${\bm b}$
and the Navier-Stokes equation (\ref{D1}) for the velocity
fluctuations ${\bm u}$ in Fourier space, we
derive an evolution equation for the cross-helicity tensor,
\begin{eqnarray*}
g_{ij}({\bm k},{\bm R},t) &=&\int \overline{b_i({\bm k}
+ {\bm  K} / 2,t) \,  u_j(-{\bm k} + {\bm  K} / 2,t)}
\, \exp[i {\bm K} {\bm \cdot} {\bm R}] \,d {\bm  K} ,
\end{eqnarray*}
which depends on the velocity correlation function:
\begin{eqnarray*}
f_{ij}({\bm k},{\bm R},t) = \int \overline{u_i({\bm k}
+ {\bm  K} / 2,t) \,  u_j(-{\bm k} + {\bm  K} / 2,t)}
\, \exp[i {\bm K} {\bm \cdot} {\bm R}] \,d {\bm  K} .
\end{eqnarray*}
The evolution equation for the cross-helicity tensor reads:
\begin{eqnarray}
&& {\partial g_{ij}({\bm k}) \over \partial t}
= i({\bm k} {\bm \cdot} \overline{\bm B}) \, f_{ij}({\bm k})
- i k_m \overline{B}_i \, f_{mj}({\bm k})
+ I_{ij}({\bm k})+ g_{ij}^{\rm(N)}({\bm k}) ,
\label{D8}
\end{eqnarray}
where, for brevity of notation, we omit the large-scale
variables $t$ and ${\bm R}$ in the
functions $f_{ij}({\bm k},{\bm R},t)$, $g_{ij}({\bm k},{\bm R},t)$,
$g_{ij}^{\rm(N)}({\bm k},{\bm R},t)$ and $\overline{\bm B}({\bm R},t)$.
Here the term $g_{ij}^{\rm(N)}$
\begin{eqnarray}
g_{ij}^{\rm(N)}({\bm k}) &=& \int \biggl[\, \overline{b_i({\bm k}_1) \,
{\partial u_j({\bm k}_2) \over \partial t}} +
\overline{b_i^{\rm(N)}({\bm k}_1) \, u_j({\bm k}_2)}\biggr]
\, \exp[i {\bm K} {\bm \cdot} {\bm R}] \,d {\bm  K}
\label{TTT2}
\end{eqnarray}
is determined by the third-order moments appearing due to the nonlinear terms.
Note that since the Navier-Stokes equation is nonlinear,
it appears only in the third-order moments $g_{ij}^{\rm(N)}$.
Here we took into account that the fluid pressure is nonlinear in the fluctuations.
For large magnetic and fluid Reynolds numbers, the dissipative
terms caused by the kinematic viscosity $\nu$ and magnetic
diffusivity $\eta$ in (\ref{D8}) are negligibly small in comparison
with the nonlinear terms.
The term, $I_{ij}$, which contains the large-scale
spatial derivatives of both, the mean magnetic field and the turbulent
intensity, is given by
\begin{eqnarray}
I_{ij}({\bm k}) &=& {1 \over 2}\left[ (\meanBB {\bm \cdot} \bec{\nabla})\,
f_{ij} - k_{n} \, \left({\partial f_{ij} \over \partial k_{s}}\, \nabla_s \meanB_{n}
- {\partial f_{nj} \over \partial k_{s}} \, \nabla_s \meanB_{i}\right)
- \nabla_n \left(\meanB_{i} f_{nj} \right) \right] .
\label{M3}
\end{eqnarray}
This term determines the turbulent magnetic diffusivity
and effects of inhomogeneous turbulence.
The derivation of (\ref{D8}) is given in Appendix~\ref{appB}.
We consider the case of weak mean magnetic fields, i.e., the energy of the mean
magnetic field is much less than the turbulent kinetic energy.
This implies that in the present study we do not investigate quenching of the turbulent
transport coefficients.
Therefore, we do not need evolution equations for the second moments
of velocity, $\langle u_i u_j\rangle$, and magnetic field, $\langle b_i b_j\rangle$
\citep[see, e.g.,][]{rogachevskii2000,rogachevskii2001,rogachevskii2004}.
This implies that we consider linear effects in the mean magnetic field.
The nonlinear mean-field theory for the mean electromotive force in a turbulent
compressible fluid flow is the subject of a separate
ongoing study.

\subsubsection{$\tau$ approach}

Equation~(\ref{D8}) for the second moment
includes first-order spatial differential
operators $\hat{\cal M}$ applied to the
third-order moments $F^{\rm(III)}$. The problem arises
how to close (\ref{D8}), i.e., how to express
the third-order term $\hat{\cal M} F^{\rm(III)}$
through the lower moments $F^{\rm(II)}$
\citep{monin2013,mccomb1990}. We use the spectral
$\tau$ approximation which postulates that the
deviations of the third-moment terms, $\hat{\cal
M} F^{\rm(III)}({\bm k})$, from the contributions to
these terms afforded by the background
turbulence, $\hat{\cal M} F^{\rm(III,0)}({\bm k})$,
can be expressed through similar deviations
of the second moments, $F^{\rm(II)}({\bm k}) -
F^{\rm(II,0)}({\bm k})$:
\begin{eqnarray}
&& \hat{\cal M} F^{\rm(III)}({\bm k}) - \hat{\cal M}
F^{\rm(III,0)}({\bm k})
= - {1 \over \tau_r(k)} \,
\Big[F^{\rm(II)}({\bm k}) - F^{\rm(II,0)}({\bm k})\Big] ,
\label{F2}
\end{eqnarray}
where $\tau_r(k)$ is the scale-dependent relaxation time, which can be
identified with the correlation time $\tau(k)$ of the
turbulent velocity field for large fluid and magnetic Reynolds
numbers \citep{orszag1970,pouquet1976,kleeorin1990,rogachevskii2004}.
The functions with the superscript $(0)$ correspond
to the background turbulence with zero mean magnetic field.
We take into account that $g_{ij}^{(0)}=0$, because when the mean
magnetic field is zero, the electromotive force vanishes.
We do not take into account magnetic fluctuations caused by a small-scale dynamo
(i.e., a dynamo with zero mean magnetic field).
Therefore, (\ref{F2}) for $g_{ij}({\bm k})$ reduces to
$\hat{\cal M} F_i^{\rm(III)}({\bm k}) = - F_i({\bm k})
/ \tau(k)$.
Validation of the $\tau$ approximation for different
situations has been performed in various numerical
simulations and analytical studies
\citep{brandenburg2005,rogachevskii2007,rogachevskii2011,rogachevskii2012,brandenburg2012,kapyla2012}.

\subsubsection{Model of background turbulence
for ${\rm Re} \gg 1$}

In the next step of the derivation we need
a model for the background turbulence.
We use statistically stationary, density-stratified, inhomogeneous,
compressible and helical background turbulence,
which is determined by the following correlation
function in ${\bm k}$ space:
\begin{eqnarray}
f_{ij}^{(0)}({\bm k}) &=&
{E(k) \over 8 \pi \, k^2 \,  (1+ \sigma_c)}
\biggl\{ \Big[\delta_{ij} - k_{ij} + {i \over k^2} \, \big(k_j \lambda_i
- k_i \lambda_j\big) + 2\sigma_c \, k_{ij}
\nonumber\\
&& + (1 +2\sigma_c) \, {i
\over 2 k^2} \, \big(k_i \nabla_j- k_j
\nabla_i\big) \Big] \, \overline{ {\bm u}^2 }
- {i \over k^2} \, \varepsilon_{ijp} \, k_p \chi \biggr\} .
\label{B6}
\end{eqnarray}
Note that most statements and conditions used for the derivation of
the tensorial structure of (\ref{B5}) are also valid for (\ref{B6}).
We assume here that the background turbulence is of Kolmogorov type with
constant fluxes of energy and kinetic helicity over the spectrum,
i.e., the kinetic energy spectrum for the
range of wave numbers $k_0<k<k_\nu$ is
$E(k) = - d \bar \tau(k) / dk$, the function $\bar \tau(k) =
(k / k_{0})^{1-q}$ with $1 < q < 3$ being the
exponent of the kinetic energy spectrum ($q =
5/3$ for a Kolmogorov spectrum).
The condition $q>1$ corresponds to finite kinetic energy
for very large fluid Reynolds numbers, while $q<3$ corresponds to finite
dissipation of the turbulent kinetic energy at the viscous scale
\citep[see, e.g.,][]{monin2013,mccomb1990,frisch1995}.
The turbulent correlation time in ${\bm k}$ space
is $\tau(k) = 2\tau_0 \, \bar \tau(k)$,
where $\tau_0 = \ell_{0} / \sqrt{\overline{ {\bm u}^2 }}$ is the
turbulent correlation time in physical space,
and $\sqrt{\overline{ {\bm u}^2 }}$ is the
characteristic turbulent velocity at scale $\ell_{0}$.
Note that for fully developed Kolmogorov like turbulence,
$\sigma_c < 1$ \citep{chassaing2013}.

\subsubsection{Mean electromotive force
for ${\rm Rm} \gg 1$}

We use the spectral $\tau$ approach and assume
that the characteristic time of variation of the mean
magnetic field $\overline{\bm B}$ is substantially
larger than the correlation time $\tau(k)$ for all
turbulence scales. This allows us to get a stationary
solution of (\ref{D8}).
We consider linear effects in the mean magnetic field,
so that the function $f_{ij}({\bm k})$
in (\ref{D8}) and~(\ref{M3}) should be replaced by
$f_{ij}^{(0)}({\bm k})$.
The mean electromotive force is determined by
$\overline{\cal E}_{i} = \varepsilon_{inm} \, \int g_{mn}({\bm k})
\,{\rm d} {\bm k}$.
We take into account that the terms with symmetric
tensors with respect to the indexes $m$ and $n$
in $g_{mn}({\bm k})$ do not contribute to the mean
electromotive force.
Therefore, the equation for the mean electromotive force is given by
\begin{eqnarray}
\overline{\cal E}_{m} &=&  \varepsilon_{mji} \, \int \tau(k) \biggl\{i({\bm k} {\bm \cdot} \overline{\bm B}) \, f_{ij}^{(0)} - \overline{B}_i \, \left(i k_{n} \,f_{nj}^{(0)}
+ {1 \over 2} \nabla_n f_{nj}^{(0)} \right)
\nonumber\\
&& \quad \quad - \left[f_{nj}^{(0)} + {k \over 2} \, \left({{\rm d} \ln \tau \over {\rm d} k}\right)\, k_{np} \, f_{pj}^{(0)} \right] \nabla_n \meanB_{i} \biggr\} \,{\rm d} {\bm k} .
\label{DDD8}
\end{eqnarray}
Using the model of background turbulence given by (\ref{B6})
and integrating in ${\bm k}$-space (\ref{DDD8}), we arrive at equation
for the mean electromotive force~(\ref{C2}), where
the $\alpha$ effect,
the turbulent magnetic diffusivity $\etat$ and
the pumping velocity $\bm{\gamma}$
for large magnetic Reynolds numbers $({\rm Rm} \gg 1)$
are given by
\begin{eqnarray}
\alpha &=& - {1\over 3} \, {\tau_0 \, \overline{{\bm u} \,{\bf\cdot}  \,(\bec{\nabla} {\bf \times} \, {\bm u})} \over 1 + \sigma_c} ,
\label{C00}\\
\etat &=&  {\tau_0 \, \overline{{\bm u}^2} \over 3}  \,
\left[1 - {(q-1)  \, \sigma_c \over 2 (1+\sigma_c)}\right],
\label{C5}\\
\bm{\gamma} &=& - {1 \over 6} {\bm \nabla} \left(\tau_0 \,
\overline{{\bm u}^2} \right) .
\label{C6}
\end{eqnarray}
Equations~(\ref{C00}) and~(\ref{C5}) imply that for large
magnetic Reynolds numbers, compressibility decrease both
the $\alpha$ effect and turbulent magnetic diffusivity.
Since $1 < q < 3$, the turbulent magnetic diffusivity
is always positive for ${\rm Rm} \gg 1$, even for very large
compressibility, $\sigma_c \gg 1$ (e.g., for irrotational flows).
Note that for an incompressible flow
the turbulent magnetic diffusivity is independent of $q$.
On the other hand, compressibility of fluid flow
does not affect the pumping velocity $\bm{\gamma}$ of the mean
magnetic field
for ${\rm Rm} \gg 1$, similarly to the case of ${\rm Rm} \ll 1$.

In the framework of mean-field dynamo theory, the threshold for the generation of a
mean magnetic field is formulated in terms of the dynamo number.
In the case of an $\alpha^2$ dynamo, the dynamo number is given by
$R_\alpha = \alpha L_B / \etat$. Using (\ref{C00}) and~(\ref{C5}),
we find that, for a compressible flow, $R_\alpha$ is given by
\begin{eqnarray}
R_\alpha = R_\alpha^{(\rm in)} \left(1 + {3-q \over 2}\, \sigma_c\right)^{-1} ,
\label{DN20}
\end{eqnarray}
where $R_\alpha^{(\rm in)}$ applies to the corresponding incompressible flow.
This implies that compressibility decreases the dynamo number.

\section{Turbulent transport of particles}

In this section we consider non-inertial particles or gaseous admixtures
in a compressible fluid flow.
Equation for the particle number density,
$n^{\rm(p)}({\bm x},t)$, reads \citep{chandrasekhar1943,akhiezer1981}:
\begin{eqnarray}
{\partial n^{\rm(p)} \over \partial t} + \bec{\nabla} {\bm \cdot}
\, (n^{\rm(p)} \UU - D \bec{\nabla} n^{\rm(p)}) =0 ,
\label{B1}
\end{eqnarray}
where $D$ is the microphysical diffusivity describing Brownian diffusion of
particles and $\UU$ is a fluid velocity.
Equation~(\ref{B1}) implies conservation of the total
number of particles in a closed volume.
We consider one way coupling, i.e., we take into account
the effect of turbulence on particle transport, but neglect
the effect of particles on the turbulence. This corresponds to
turbulent transport of a passive scalar.

To determine the turbulent flux of particles, we use
a mean-field approach in which the number density
of particles and fluid velocity are decomposed into
mean and fluctuating parts, where the fluctuating parts
have zero mean values.
Averaging (\ref{B1}) over an ensemble, we arrive at an equation for the
mean number density, $N({\bm x},t) \equiv \overline{n^{\rm(p)}}$:
\begin{eqnarray}
{\partial N \over \partial t} + \bec\nabla {\bm \cdot} \,
\left(\overline{n \, {\bm u}} - D \bec\nabla N\right)=0 ,
\label{ME0}
\end{eqnarray}
where ${\bm F}=\overline{n({\bm x},t) \, {\bm u}({\bm x},t)}$ is the turbulent
flux of particles and we consider the case of a zero mean fluid velocity,
$\overline{\bm U}=0$.

To determine the turbulent flux of particles we use
the equation for the fluctuations of the particle number density,
$n({\bm x},t)=n^{\rm(p)}-N$, which follows from (\ref{B1})
and~(\ref{ME0}):
\begin{eqnarray}
{\partial n \over \partial t} + \bec\nabla {\bm \cdot}
\, \left(n \, {\bm u} - \overline{n \, {\bm u}}
- D \bec\nabla n\right)  = -N \, \bec\nabla {\bm \cdot} \, {\bm u}
-  ({\bm u}{\bm \cdot} \bec{\nabla}) N,
\label{E1}
\end{eqnarray}
where ${\cal Q}=\bec\nabla {\bm \cdot}
\, (n {\bm u} - \overline{n {\bm u}})$ are nonlinear terms and $I = -N \,
\bec\nabla {\bm \cdot} \, {\bm u} -  ({\bm u}
{\bm \cdot} \bec{\nabla}) N$ is the source term
of particle number density fluctuations.
The ratio of the nonlinear terms to the diffusion
term is the P\'eclet number, that is estimated as
${\rm Pe} = u_{0} \, \ell_0 / D$.
In the next subsections we derive equations for the
turbulent transport coefficients of particles for
small and large P\'eclet numbers.

\subsection{Turbulent transport coefficients for ${\rm Pe} \ll 1$}

We derive an equation for the turbulent flux of particles,
\begin{eqnarray}
\overline{n({\bm x},t) \, u_j ({\bm  x},t)} = \int F_{j}({\bm k},\omega,{\bm R},t)
\,d\omega \,d {\bm k} ,
\label{EEE5}
\end{eqnarray}
for small P\'eclet numbers using a quasi-linear approach, where
\begin{eqnarray}
F_{j}({\bm k},\omega,{\bm R},t) =
 \int \overline{n({\bm k}_1,\omega_1) \, u_j({\bm k}_2,\omega_2)}
\exp[i \Omega t+ i {\bm K} {\bm \cdot} {\bm R}] \,d \Omega \,d {\bm  K}.
\label{EEE6}
\end{eqnarray}
For brevity of notations we omit below the large-scale variables $t$ and ${\bm R}$
in $F_{j}({\bm k},\omega,{\bm R},t)$ and $N({\bm R},t)$.
We neglect the nonlinear term ${\bec {\cal Q}}$, but keep
the molecular diffusion term in (\ref{E1}).
We rewrite (\ref{E1}) in Fourier space and find
the solution of this equation (see (\ref{EG1}) in Appendix~\ref{appA}).
Here we apply the same approach that was used in
section~\ref{small-Rm}, i.e.,
we assume that there is a separation of spatial and time scales,
$\ell_0 \ll L_N$ and $\tau_0 \ll t_N$, where
$L_N$ and $t_N$ are the characteristic spatial and time scales
of the mean particle number density variations.
We perform calculations presented in Appendix~\ref{appA}, so that equation for
the turbulent flux of particles is given by
\begin{eqnarray}
F_j &=& - \int G_D \biggl[N \left(i k_i f_{ij} + {1 \over 2} \nabla_i f_{ij} - D k^2 G_D \, k_{im} \nabla_m f_{ij}\right) + {1 \over 2} \biggl(f_{mj} - k_i
{\partial f_{ij} \over \partial k_m}
\nonumber\\
&&- 2 D k^2 G_D \, k_{im} f_{ij} \biggr)\nabla_m N \biggr] \,d {\bm k} \,d\omega ,
\label{E5}
\end{eqnarray}
where $G_D \equiv G_D({\bm k},\omega) = (D {\bm k}^2 + i \omega)^{-1}$ and
$f_{ij}=f_{ij}({\bm k},\omega,{\bm R},t)$.
Since we consider one way coupling, the correlation function
$f_{ij}$ in (\ref{E5}) should be replaced by $f_{ij}^{(0)}$
for the background random flow with zero turbulent flux of particles.
Using the model of background random flow given by (\ref{B5})
with zero kinetic helicity, and
integrating in $\omega$ and ${\bm k}$-space, we arrive at an equation
for the turbulent flux of particles:
\begin{eqnarray}
{\bm F} = N \, {\bm V}^{\rm eff} - \Dt \, \bec\nabla N ,
\label{B7}
\end{eqnarray}
where for small P\'eclet numbers $({\rm Pe} \ll 1)$, the
turbulent diffusivity $\Dt$ and the effective pumping
velocity ${\bm V}^{\rm eff}$ are given by
\begin{eqnarray}
\Dt &=&  {(q-1)\over 3(q+1)} \, \tau_0 \, \overline{{\bm u}^2} \, \left(1 - {2 \sigma_c
\over 1 + \sigma_c} \right) \, {\rm Pe}
\equiv {(q-1)\over 3(q+1)} \, \left({1- \sigma_c
\over 1 + \sigma_c} \right) \, {\rm Pe}^2 \, D ,
\label{B8}\\
{\bm V}^{\rm eff} &=& {(q-1)\over 3(q+1)} \, \tau_0 \, \overline{{\bm u}^2}  \,
\left[{1 \over (1 + \sigma_c)} {\bm \nabla} \ln \meanrho + {3 \sigma_c
\over 2(1 + \sigma_c)} {\bm \nabla} \ln \overline{{\bm u}^2} \right] \, {\rm Pe} .
\label{BB9}
\end{eqnarray}
Since $\tau_0 \, {\rm Pe} = \ell_0^2 /D$, the turbulent transport coefficients
given by (\ref{B8}) and (\ref{BB9}) are determined only by the microphysical
diffusion time scale, $\ell_0^2 /D$.
Remarkably, equation~(\ref{B8}) for the turbulent diffusivity of passive scalars $\Dt$
coincides with (\ref{C3}) for the turbulent magnetic diffusivity $\etat$
after replacing ${\rm Pe}$ by ${\rm Rm}$.

Equation~(\ref{B8}) implies that for small
P\'eclet numbers, compressibility effects decrease
the turbulent diffusivity.
However, the total (effective) diffusivity, $D+\Dt$, cannot be negative,
because for ${\rm Pe} \ll 1$, the molecular diffusivity
is much larger than the turbulent one, $D \gg |\Dt|$.
This result is consistent with that of \cite{radler2011},
where it has been demonstrated that the total
mean-field diffusivity for passive scalar transport in irrotational flows
is smaller than the molecular diffusivity
(see section~\ref{comparison} for a detailed comparison).

The physics related to different terms in (\ref{BB9})
will be discussed in the next section.
In the present study of turbulent transport of passive scalar
and non-inertial particles, we consider nonhelical turbulence, because
there is no effect of the kinetic helicity on the particle flux---at least
in the system investigated here [without rotation, the large-scale shear,
and neglecting effects $\sim$O$(\lambda^2 \, \overline{{\bm u}^2}, \nabla^2 \, \overline{{\bm u}^2}, \lambda_i \nabla_j\overline{{\bm u}^2})$].

By means of the equation of state for a perfect gas,
${\bm \nabla} \ln \meanp={\bm \nabla} \ln \meanrho+{\bm \nabla} \ln \meanT$,
we rewrite (\ref{BB9}) for the effective pumping velocity as:
\begin{eqnarray}
{\bm V}^{\rm eff} &=& {(q-1)\over 3 (q+1)} \, \tau_0 \, \overline{{\bm u}^2}  \,
\biggl[{1 \over (1 + \sigma_c)} \biggl({\bm \nabla} \ln \meanp - {\bm \nabla} \ln \meanT\biggr) + {3 \sigma_c \over 2(1 + \sigma_c)} {\bm \nabla} \ln \overline{{\bm u}^2}
\biggr] \, {\rm Pe} ,
\label{B9}
\end{eqnarray}
where $\meanT$ and $\meanp$ are the mean temperature and the mean pressure, respectively.
Note that, since the density-temperature correlation
$\overline{\rho' \vartheta}$ is much smaller than $\meanrho \, \meanT$
\citep{chassaing2013}, the equation of state for perfect gas is
also valid for the mean quantities, where $\rho'$ and $\vartheta$ are
fluctuations of the fluid density and temperature.

\subsection{Turbulent transport coefficients for ${\rm Pe} \gg 1$}

In this section we determine the turbulent flux of particles
for large P\'eclet and Reynolds numbers.
Similar to the study performed in section~\ref{large-Rm},
we consider fully developed turbulence, where the Strouhal number
is of the order of unity and the turbulent correlation time
is scale-dependent, so we apply the Fourier transformation only in
${\bm k}$ space.
Using (\ref{E1}) for the fluctuations $n$ and the Navier-Stokes equation
for the velocity ${\bm u}$ written in Fourier space, we derive
an equation for the instantaneous two-point correlation function
\begin{eqnarray*}
F_{j}({\bm k}) &=&\int \overline{n({\bm k}
+ {\bm  K} / 2) \,  u_j(-{\bm k} + {\bm  K} / 2)}
\, \exp[i {\bm K} {\bm \cdot} {\bm R}] \,d {\bm  K} .
\end{eqnarray*}
For brevity of notations we omit the large-scale variables $t$ and ${\bm R}$
in the function $F_{j}({\bm k},{\bm R},t)$ and the mean number density $N({\bm R},t)$.
To derive an evolution equation for $F_{j}({\bm k})$, we perform calculations
presented in Appendix~\ref{appB}, which yield
\begin{eqnarray}
{\partial F_j \over \partial t} = - N \left(i k_i f_{ij}+ {1 \over 2} \nabla_i f_{ij} \right)
- {1 \over 2} \left(f_{ij} - k_{m} \, {\partial f_{mj} \over \partial k_{i}}
\right) \nabla_i N + \hat{\cal M} F_i^{\rm(III)}({\bm k}) ,
\label{T1}
\end{eqnarray}
where
\begin{eqnarray}
\hat{\cal M} F_j^{\rm(III)}({\bm k}) &=& \int \biggl[\, \overline{n({\bm k}_1) \,
{\partial u_j({\bm k}_2) \over \partial t}} -
\overline{{\cal Q}({\bm k}_1) u_j({\bm k}_2)}\biggr]
\exp[i {\bm K} {\bm \cdot} {\bm R}] \,d {\bm  K}
\label{T2}
\end{eqnarray}
are the third-order moment terms in ${\bm k}$ space appearing due
to the nonlinear terms.

We use the spectral $\tau$ approximation~(\ref{F2}), where
functions with superscript $(0)$ correspond to background
turbulence with zero turbulent particle flux.
Therefore, (\ref{F2}) reduces to
$\hat{\cal M} F_i^{\rm(III)}({\bm k}) = - F_i({\bm k})/ \tau(k)$.
We also assume that the characteristic time of variation
of the second moment $F_i({\bm k})$ is substantially
larger than the correlation time $\tau(k)$ on all
turbulence scales.
Therefore, the particle flux is given by
\begin{eqnarray}
F_j({\bm k}) = - \tau(k) \, \left\{ \left(i k_i f_{ij}^{(0)}
+ {1 \over 2} \nabla_i f_{ij}^{(0)} \right) N
+ \left[f_{mj}^{(0)} + {k \over 2} \, \left({{\rm d} \ln \tau \over {\rm d} k}\right)
\, k_{im} \, f_{ij}^{(0)} \right] \nabla_m N \right\} .
\nonumber\\
\label{TTT1}
\end{eqnarray}
Using the model of background turbulence for $f_{ij}^{(0)}$ given by (\ref{B6})
with a zero kinetic helicity, and integrating (\ref{TTT1}) in ${\bm k}$-space,
we obtain the turbulent flux of particles~(\ref{B7}), where the
turbulent diffusivity $\Dt$ and the effective pumping
velocity ${\bm V}^{\rm eff}$ of non-inertial particles
for ${\rm Pe} \gg 1$ are given by
\begin{eqnarray}
\Dt &=&  {\tau_0 \, \overline{{\bm u}^2} \over 3}  \, \left[1 - {(q-1) \sigma_c
\over 2(1 + \sigma_c)} \right] ,
\label{T3}\\
{\bm V}^{\rm eff} &=& {\tau_0 \, \overline{{\bm u}^2} \over 3} \,
\left[{1 \over (1 + \sigma_c)} {\bm \nabla} \ln \meanrho + {\sigma_c
\over 2(1 + \sigma_c)} {\bm \nabla} \ln \overline{{\bm u}^2} \right] .
\label{T4}
\end{eqnarray}
Equation~(\ref{T3}) for the turbulent diffusivity of particles $\Dt$
coincides with (\ref{C5}) for the turbulent magnetic diffusivity $\etat$
after replacing ${\rm Pe}$ by ${\rm Rm}$.
Equation~(\ref{T3}) implies that for large P\'eclet numbers, compressibility
decreases the turbulent diffusivity of particles.
Since $1 < q < 3$, the turbulent diffusivity is always positive when
${\rm Pe} \gg 1$, even for very strong compressibility, $\sigma_c \gg 1$.

\begin{figure}[t!]\begin{center}
\includegraphics[width=.7\textwidth]{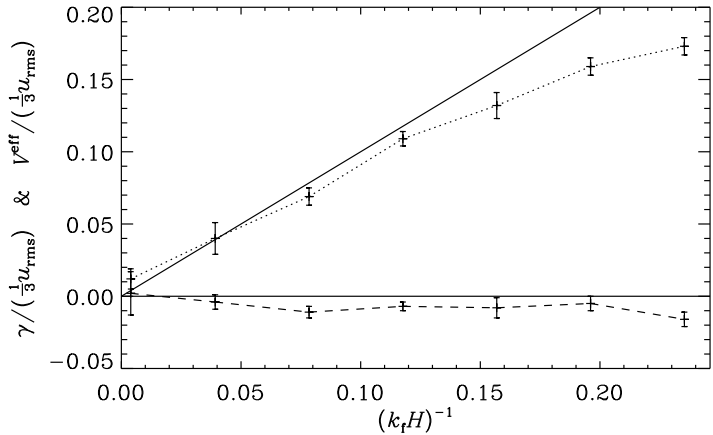}
\end{center}\caption[]{
Dependence of $V^{\rm eff}$ (dotted line) and $\gamma$ (dashed line)
(here in the $z$ direction) as a function of the inverse scale height $H$,
normalized by the wavenumber $k_{\rm f}$ of the energy-carrying eddies.
The turbulent pumping velocity $\gamma$ caused by the nonuniformity of the
turbulent rms velocity (solid line) in a nonstratified turbulence.
Here the symbol $H$ is used either for the
density scale height, $H=\lambda^{-1}$, or for the characteristic scale
of the nonuniformity of the turbulent rms velocity,
$H^{-1} = {\bm \nabla} u_{\rm rms} / u_{\rm rms}$.
}\label{psummary_gdep}\end{figure}

Equation~(\ref{T4}) determines effective pumping velocity, ${\bm V}^{\rm eff}$,
of passive scalars and non-inertial particles.
The first term in (\ref{T4}), and likewise in (\ref{BB9}) for
${\rm Pe} \ll 1$, describes pumping of passive scalars
caused by density stratification.
This effect for low Mach numbers
has been studied theoretically using various analytical
approaches \citep{elperin1995,elperin1996,elperin1997,elperin2000,
elperin2001,pandya2002,reeks2005,amir2017}.
It has also been detected in the direct numerical simulations of
\cite{brandenburg2012b}, where the density stratification was caused by
gravity.
Their result is reproduced in \Fig{psummary_gdep} and compared with
the corresponding turbulent pumping velocities of mean magnetic field,
which is close to zero.
This demonstrates very clearly the different natures of pumping
of particles and magnetic fields in one and the same simulation.
Here, the relevant components of
the effective pumping velocities of particles,
${\bm V}^{\rm eff}$, and magnetic field, $\bm{\gamma}$,
have been determined using the test-field method for axisymmetric
turbulence described in \cite{brandenburg2012b}.
The simulations have been carried out for about 200 turnover times,
for ${\rm Pe} = {\rm Re} = u_{\rm rms} / \nu k_{\rm f} = 22$
and $k_{\rm f}/k_1=5$, where $k_1$ is the wavenumber based on the size of the
computational domain
and $k_{\rm f}$ is the forcing scale of turbulence.
Error bars have been determined using any one third of the full time
series of the instantaneous, but spatially averaged mean values of
${\bm V}^{\rm eff}$ and $\bm{\gamma}$.
Within error bars, the result for $\bm{\gamma}$ is not quite compatible
with zero, but this is entirely explicable as a consequence of a small
upward increase in the rms velocity with height and thus a small
positive gradient.
Also, the deviation of the theoretical from the numerical results for
larger $(k_{\rm f} H)^{-1}$ is caused by the fact that the parameter
$\ell_0/H$ is no longer very small, as was assumed in the theory.
The effective pumping velocity of particles ${\bm V}^{\rm eff}$ 
for arbitrary stratifications has been determined analytically 
by \cite{amir2017}. 
They have shown that there is a quenching of the pumping velocity
caused by the strong stratification. This tendency is already seen
in \Fig{psummary_gdep} (dotted line) for larger density stratifications.

Let us discuss the physics related to the first term in the right hand side of (\ref{T4})
caused by the fluid density stratification, considering
density stratified homogeneous turbulence.
Substituting (\ref{B7}) into (\ref{ME0}), we obtain the equation for the mean number density, $N$:
\begin{eqnarray}
{\partial N \over \partial t} + \bec\nabla {\bm \cdot} \,
\big[N \, {\bm V}^{\rm eff} - (D+\Dt) \, \bec\nabla N \big] =0 .
\label{BBB7}
\end{eqnarray}
The steady-state solution of Eqs.~(\ref{BBB7})
for the mean number density of non-inertial particles is given by
$N/N_0 = [\meanrho/\meanrho_0]^{\Dt / (D+\Dt)}$,
where subscripts $0$ represent the values
in the region with homogeneous fluid density.
Here we used Eqs.~(\ref{B8}) and (\ref{BB9}) or (\ref{T3}) and (\ref{T4})
for homogeneous turbulence.
This solution implies that small particles
are accumulated in the vicinity of the maximum of the fluid
density,
because the non-diffusive flux of particles, $N \, {\bm V}^{\rm eff}$,
is directed toward the maximum of the fluid density.

The accumulation of non-inertial particles in the vicinity of the
maximum of the mean fluid density can be explained
as follows \citep{elperin1995,elperin1997,haugen2012}.
Let us assume that the mean fluid density is inhomogeneous
along the $x$ axis, and the mean density
$\meanrho_2$ at point $2$ is larger than the mean
density $\meanrho_1$ at point $1$. Consider
two small control volumes {``\sf{a}''} and
{``\sf{b}''} located between these two points,
and let the direction of the local turbulent
velocity in volume {``\sf{a}''} at some instant
be the same as the direction of the mean fluid
density gradient $\bec\nabla \, \meanrho$ (e.g.,
along the $x$ axis toward point $2$). Let the
local turbulent velocity in volume {``\sf{b}''}
at this instant be directed opposite to the mean
fluid density gradient (i.e., toward point $1$).
In a fluid flow with an imposed mean fluid density
gradient, one of the sources of particle number
density fluctuations, $n \propto - \tau_0 \, N \,
(\bec\nabla {\bf \cdot} \, {\bm u})$, is caused by a
non-zero $\bec\nabla\cdot{\bm u}\approx - {\bm u} \cdot
\bec\nabla \ln \meanrho \not=0$ [see the first term
on the right hand side of Eq.~(\ref{E1})].
Since fluctuations of the fluid velocity $u_x$ are
positive in volume {``\sf{a}''} and negative in
volume {``\sf{b}''}, $\bec\nabla {\bf
\cdot} \, {\bm u} < 0$ in volume {``\sf{a}''}, and
$\bec\nabla {\bf \cdot} \, {\bm u} > 0$ in volume
{``\sf{b}''}.
Therefore, the fluctuations of the
particle number density $n \propto - \tau_0 \, N
\, (\bec\nabla {\bf \cdot} \, {\bm u})$ are positive
in volume {``\sf{a}''} and negative in volume
{``\sf{b}''}. However, the flux of particles $n\,
u_x$ is positive in volume {``\sf{a}''} (i.e., it
is directed toward point $2$), and it is also
positive in volume {``\sf{b}''} (because both
fluctuations of fluid velocity and number density
of particles are negative in volume
{``\sf{b}''}). Therefore, the mean flux of
particles $\overline{n\, {\bm u}}$ is directed, as is
the mean fluid density gradient $\bec\nabla \,
\meanrho$, toward point~2. This forms large-scale
heterogeneous structures of non-inertial
particles in regions with a mean fluid density
maximum.

Indirect manifestations of the pumping effect
caused by the density stratification through an observed
increase of small-scale clustering of particles can be identified
in the direct numerical simulations of \cite{clercx2008}, who studied
the evolution of inertial particles in stably stratified turbulence.
This pumping effect increases the small-scale clustering of inertial particles
by forming inhomogeneous distributions of the mean particle number density.
Note also that tangling of the mean particle number density, $N$, by
compressible velocity fluctuations (described by the source term
$-N {\rm div} \, {\bm u} $ in (\ref{E1})),
can increase the level of particle number density fluctuations, $n$,
when the source term $-N {\rm div} \, {\bm u} > 0$.
In this case $\partial n / \partial t > 0$, see (\ref{E1}).
This can amplify the particle clustering \citep{eidelman2010,elperin2013}.

Using the equation of state for a perfect gas,
we rewrite (\ref{T4}) for the pumping effective velocity in the form
\begin{eqnarray}
{\bm V}^{\rm eff} &=& {\tau_0 \, \overline{{\bm u}^2} \over 3} \,
\biggl[{1 \over (1 + \sigma_c)} \biggl({\bm \nabla} \ln \meanp - {\bm \nabla} \ln \meanT\biggr) + {\sigma_c \over 2(1 + \sigma_c)} {\bm \nabla} \ln \overline{{\bm u}^2}
\biggr] .
\label{T5}
\end{eqnarray}
The first term in the right hand side of (\ref{T5})
describes turbulent barodiffusion \citep{elperin1997},
while the second term characterizes the phenomenon of turbulent
thermal diffusion \citep{elperin1996,elperin2000,amir2017}.
The last term in (\ref{T5}), $\propto \sigma_c {\bm \nabla} \overline{{\bm u}^2}$,
describes the compressible turbophoresis for noninertial particles.

The classical turbophoresis
\citep{caporaloni1975,reeks1983,reeks1992,elperin1998,mitra2018}
exists only for inertial particles and causes an additional mean particle
velocity that is proportional to
$- \tau_s {\bm \nabla} \overline{{\bm u}^2}$, where
$\tau_s= m_p / 6 \pi \rho \, \nu a_p$ is the particle Stokes time, i.e.,
the characteristic time of coupling between
small spherical particles (of radius $a_p$ and mass $m_p$) and surrounding fluid.
To illustrate the classical turbophoresis effect,
let us consider small inertial particles
in an isotropic incompressible turbulence
and use equation of motion for a particle:
$d{\bm v}^{\rm(p)}/dt = - ({\bm v}^{\rm(p)}-{\bm u})/\tau_s$,
where ${\bm v}^{\rm(p)}$ is the particle velocity.
This equation represents a balance of particle inertia with the
fluid drag force produced by the motion of the particle
relative to the surrounding fluid. Solution of this equation
for small Stokes time is given by:
${\bm v}^{\rm(p)} = {\bm u} - \tau_s [\partial {\bm u} /
\partial t + ({\bm u} \cdot {\bm \nabla}) {\bm u}] +
{\rm O}(\tau_s^2)$ \citep{maxey1987}.
Averaging this equation over statistics of fluid velocity fluctuations,
we determine the mean particle velocity:
$\overline{v}^{\rm(p)}_j = - \tau_s \nabla_j \overline{u_i u_j}$.
We take into account that for isotropic turbulence
$\overline{u_i u_j}= (1/3)\overline{{\bm u}^2} \delta_{ij}$.
Therefore, the mean particle velocity is
$\overline{\bm v}^{\rm(p)} = - (\tau_s/3) {\bm \nabla} \overline{{\bm u}^2}$
\citep[for details, see][]{elperin1998}.
This turbophoresis effect results in the formation of large-scale
clusters of inertial particles
at local minima of the mean-squared turbulent velocity.
Note that the compressible turbophoresis
for noninertial particles originates from the turbulent particle flux
$\overline{n \, {\bm u}}$, i.e., it describes
a collective statistical phenomenon, while
classical turbophoresis is not related to the
correlations between velocity and number density
fluctuations and is obtained from the expression for the mean
velocity of inertial particles.

In the case of rapidly rotating turbulence,
another pumping effect of inertial particles arises
which can accumulate particles \citep{hodgson1998,elperin1998b}.
This effect can be related to the combined action of
particle inertia and non-zero mean kinetic helicity
in rotating turbulence.

\section{Comparison with \cite{krause1980} and \cite{radler2011}}
\label{comparison}

In this section we compare our results with those obtained previously
by \cite{krause1980} and \cite{radler2011}.
Let us consider non-stratified, isotropic, homogeneous and non-helical random flow.
In this case (\ref{B5}) reads:
\begin{eqnarray}
f_{ij}^{(0)}({\bm k},\omega,{\bm R}) &=&
{\overline{{\bm u}^2} \, \Phi_u(\omega) \, E(k) \over 8 \pi \, k^2 \,
(1+ \sigma_c)} \, \Big[\delta_{ij} - k_{ij} + 2 \sigma_c \, k_{ij}  \Big]  ,
\label{B56}
\end{eqnarray}
where the first two terms, $\delta_{ij} - k_{ij}$, in the squared brackets
describes the vortical part
of the turbulent flow, while the last term, $2 \sigma_c \, k_{ij}$
determines the potential part of the flow.
Representing the velocity fluctuations as the sum of vortical and potential parts,
\begin{eqnarray}
{\bm u}=\bec{\nabla} {\bm \times} \bec{\psi} + \bec{\nabla} \phi,
\label{S120}
\end{eqnarray}
we can construct the correlation function, $f_{ij}^{(0)}({\bm k})=\overline{u_i({\bm k}) \, u_j (-{\bm  k})}$, of the velocity field in ${\bm k}$ space
for non-stratified, isotropic, homogeneous and non-helical random background flow as
\begin{eqnarray}
f_{ij}^{(0)}({\bm k}) = {\overline{{\bm u}^2} \, E(k) \over 8 \pi k^2 \,
\left(1 + \overline{\phi^2}/\overline{\bec{\psi}^2} \, \right)} \, \left(\delta_{ij} - k_{ij} + {2 \, \overline{\phi^2} \over \overline{\bec{\psi}^2}} \, k_{ij}\right),
\label{S11}
\end{eqnarray}
where $\bec{\nabla} {\bm \cdot} \bec{\psi}=0$, and we also assumed that
compressible and incompressible parts of the random velocity field have
the same spectra, $E(k)$.
It follows from (\ref{B56}) and~(\ref{S11}) that the degree of compressibility
is $\sigma_c = \overline{\phi^2}/\overline{\bec{\psi}^2}$.
For simplicity, we consider here the instantaneous correlation function.

To derive (\ref{S11}), we rewrite (\ref{S120}) in ${\bm k}$ space:
$u_i({\bm k}) = ik_m \varepsilon_{imn} \psi_n({\bm k}) + i k_i \phi({\bm k})$,
and assume that the vortical and potential parts 
are not correlated.
This allows us to obtain the following expression for $f_{ij}^{(0)}({\bm k})$:
\begin{eqnarray}
f_{ij}^{(0)}({\bm k}) = k^2 \Big[\varepsilon_{imn} \varepsilon_{jpq} \, k_{mp}
\, \overline{\psi_n({\bm k}) \, \psi_q (-{\bm  k})} + k_{ij} \, \overline{\phi({\bm k}) \, \phi(-{\bm  k})} \Big],
\label{S121}
\end{eqnarray}
where
\begin{eqnarray}
\overline{\bec{\psi}^2}=\int \overline{\psi_n({\bm k}) \, \psi_n (-{\bm  k})} \,d{\bm  k}, \quad \overline{\phi^2}=\int \overline{\phi({\bm k}) \, \phi(-{\bm  k})} \,d{\bm  k}.
\label{S124}
\end{eqnarray}
Taking into account that for the vortical field $\bec{\nabla} {\bm \cdot} \bec{\psi}=0$,
the correlation function is of the form
$\overline{\psi_i({\bm k}) \, \psi_j (-{\bm  k})} \propto \delta_{ij} - k_{ij}$,
and using (\ref{S124}) we obtain
\begin{eqnarray}
&& \overline{\psi_i({\bm k}) \, \psi_j (-{\bm  k})} = {\overline{{\bm u}^2} \, E(k) \over 8 \pi k^4 \, \left(1 + \overline{\phi^2} /\overline{\bec{\psi}^2}\right)} \,
\left(\delta_{ij} - k_{ij}\right) ,
\label{S122}\\
&& \overline{\phi({\bm k}) \, \phi(-{\bm  k})} = {\overline{{\bm u}^2} \, E(k) \over 4 \pi k^4 \, \left(1 + \overline{\bec{\psi}^2}/ \overline{\phi^2} \right)} .
\label{S123}
\end{eqnarray}
Substituting (\ref{S122})-(\ref{S123}) into (\ref{S121}), we arrive at (\ref{S11}).

\subsection{Small magnetic Reynolds numbers}

Substituting (\ref{S11})
into (\ref{G3}) and using (\ref{G2}), we find
that the turbulent magnetic diffusivity for small magnetic Reynolds numbers
reads:
\begin{eqnarray}
\etat= {1 \over 3 \eta} \, \left(\overline{\bec{\psi}^2} - \overline{\phi^2} \right).
\label{S111}
\end{eqnarray}
This result agrees with that of \cite{krause1980}.
Substituting (\ref{S11}) into (\ref{E5}), we find
that the turbulent diffusion coefficient for passive scalars
at small P\'eclet numbers is given by:
\begin{eqnarray}
\Dt= {1 \over 3 D} \, \left(\overline{\bec{\psi}^2} - \overline{\phi^2} \right).
\label{S112}
\end{eqnarray}
This result also agrees with that of \cite{radler2011}.

\subsection{Large magnetic Reynolds numbers}

In the case of large magnetic Reynolds numbers, substituting (\ref{S11})
into (\ref{DDD8}), we find for the turbulent magnetic diffusivity
\begin{eqnarray}
\etat= {\tau_0 \, \overline{{\bm u}^2} \over 3} \, \left[1 - {(q-1) \over 2} \, \left({ \overline{\phi^2} \over \overline{\bec{\psi}^2} + \overline{\phi^2}} \right) \right] .
\label{S114}
\end{eqnarray}
This result is in good agreement with (\ref{C5}) for $\etat$
and with (\ref{T3}) for $\Dt$.

\bigskip

\section{Conclusions}

A mean-field theory for the compressibility effects in the
mean electromotive force in turbulent magnetohydrodynamics and
in the turbulent flux of passive scalars or particles has been developed.
This study is based on the quasi-linear approach
applied for small magnetic Reynolds and P\'eclet numbers,
and on the spectral $\tau$-approach
for large fluid (Re) and magnetic (Rm) Reynolds numbers
and large P\'eclet numbers (Pe).
Compressibility is found to cause a depletion of
the $\alpha$ effect and turbulent magnetic diffusion
for small and large Rm.
It also decreases the turbulent diffusivity of passive scalars
for small and large Pe.
Indeed, the expressions for the turbulent magnetic diffusivity
and the turbulent diffusivity of passive scalars coincide
after replacing Rm by Pe, and vise versa.
Compressibility does not change the effective
pumping velocity of the magnetic field for Rm $\gg 1$, but decreases it
for Rm $\ll 1$.
In addition, compressibility
causes a pumping of particles in regions with higher turbulent intensity
for small and large P\'eclet numbers.
This new effect is interpreted in terms of compressible turbophoresis
for noninertial particles.
This effect is completely different from classical
turbophoresis, which only affects inertial particles and results in
pumping of inertial particles to regions of lower
turbulent intensity.
Compressible turbophoresis for noninertial particles
is a collective statistical phenomenon
originating from the turbulent particle flux.
On the other hand, classical turbophoresis
originates directly from the expression for the mean
velocity of inertial particles.
For small Mach numbers, compressibility effects are only
determined by the density stratification
through the $\lambda$ terms in (\ref{B5}) and (\ref{B6}).
We find that the density stratification does not affect
the magnetic field, although it causes a turbulent pumping
of particles to regions of maximum mean
fluid density.

The mean-field theory developed for compressibility effects in
turbulent magnetohydrodynamics and turbulent transport of passive
scalars and particles needs to be verified in direct numerical simulations
and large-eddy simulations using the test-field and test-scalar methods
\citep[see, e.g.,][]{schrinner2005,schrinner2007,brandenburg2008,brandenburg2012b},
which is the subject of a separate study.
Furthermore, in view of astrophysical and geophysical applications, it is important
to consider the additional effects of rotation, turbulent convection,
and the presence of stably stratified turbulence
with large-scale temperature gradients
\citep[see, e.g.,][]{kleeorin2003,rogachevskii2006,rogachevskii2007,brandenburg2012b}.

\begin{acknowledgements}
We thank Matthias Rheinhardt for his many questions and suggestions
which have significantly improved the clarity of the presentation.
This work was supported in part by the Research Council of Norway
under the FRINATEK (grant No.\ 231444),
the Israel Science Foundation governed by the Israeli
Academy of Sciences (grant No. 1210/15),
the National Science Foundation under grants No.\ NSF PHY-1748958
and AAG-1615100,
and the University of Colorado through its support of the
George Ellery Hale visiting faculty appointment.
I.R. acknowledges the hospitality of Nordita,
the Laboratory for Atmospheric and Space Physics of the University of Colorado,
the Kavli Institute for Theoretical Physics in Santa Barbara
and the \'Ecole Polytechnique F\'ed\'erale de Lausanne.
\end{acknowledgements}

\appendix
\section{Derivation of (\ref{G2}) for Rm $\ll 1$, and (\ref{E5}) for Pe $\ll 1$}
\label{appA}

In the limit of small magnetic Reynolds number, we neglect in (\ref{D2})
nonlinear terms, but keep molecular dissipative
terms of the magnetic fluctuations.
The solution of (\ref{D2}) reads:
\begin{eqnarray}
&& b_i({\bm k}, \omega) = i k_p G_\eta({\bm k}, \omega) \,
\biggl[\int \meanB_p({\bm Q}) u_i({\bm k} - {\bm Q}, \omega) \,d{\bm Q}
- \int \meanB_i({\bm Q}) \, u_p({\bm k} - {\bm Q}, \omega) \,d{\bm Q} \biggr],
\nonumber\\
 \label{G1}
\end{eqnarray}
where $G_\eta({\bm k},\omega) = (\eta {\bm k}^2 + i \omega)^{-1}$.
To derive (\ref{G1}), we consider the source term in the induction equation
for the magnetic fluctuations, $(\overline{\bm B} {\bf \cdot} \bec{\nabla}){\bm u}$,
which can be rewritten using Fourier transformation as
\begin{eqnarray}
(\overline{\bm B} {\bf \cdot} \bec{\nabla}) u_i &=&
\int \overline{B}_p({\bm Q}) \, \exp (i {\bm Q} {\bm \cdot} {\bm x}) \,d{\bm Q} \,
\int i k'_p \, u_i({\bm k}') \, \exp (i {\bm k}' {\bm \cdot} {\bm x})
\,d{\bm k}' .
\label{R20}
\end{eqnarray}
For brevity of notation
we omit hereafter the arguments $t$ and ${\bm x}$ in the functions $\overline{\bm B}({\bm x},t)$ and ${\bm u}({\bm x},t)$, and also the variable $t$ in the functions
$\overline{B}_p({\bm Q},t)$ and $u_i({\bm k},t)$.
We introduce a new variable ${\bm k}={\bm k}'+{\bm Q}$ and take into account
that $Q_p \, \overline{B}_p({\bm Q}) = 0$, which yields
\begin{eqnarray}
(\overline{\bm B} {\bf \cdot} \bec{\nabla}) u_i
&=& \int \left(i k_p \, \int \overline{B}_p({\bm Q}) \, u_i({\bm k}-{\bm Q}) \,d{\bm Q} \right) \, \exp (i {\bm k} {\bm \cdot} {\bm x}) \,d{\bm k}
\nonumber\\
&\equiv& \int \left[(\overline{\bm B} {\bf \cdot} \bec{\nabla}) u_i\right]_{\bm k} \exp (i {\bm k} {\bm \cdot} {\bm x}) \,d{\bm k} ,
\label{R21}
\end{eqnarray}
where
$[...]_{\bm k}$ denotes the Fourier transform.
Therefore,
\begin{eqnarray}
\left[(\overline{\bm B} {\bf \cdot} \bec{\nabla}) u_i\right]_{\bm k} &=&
i k_p \, \int \overline{B}_p({\bm Q}) \, u_i({\bm k}-{\bm Q}) \,d{\bm Q} .
\label{R22}
\end{eqnarray}
Similar derivations are used for the other terms in the induction equation:
\begin{eqnarray}
\left[({\bm u} {\bm \cdot} \bec{\nabla}) \overline{B}_i\right]_{\bm k} &=&
i \int Q_p \, \overline{B}_i({\bm Q}) \, u_p({\bm k}-{\bm Q})  \,d{\bm Q} ,
\label{R222}\\
\left[\overline{B}_i \, (\bec{\nabla} {\bm \cdot} {\bm u}) \right]_{\bm k} &=&
i \int \left(k_p -Q_p\right) \, \overline{B}_i({\bm Q}) \, u_p({\bm k}-{\bm Q}) \,d{\bm Q} ,
\label{R223}
\end{eqnarray}
which yield (\ref{G1}).

To derive (\ref{G2}), we use (\ref{E3}) and (\ref{G1}) which yield
\begin{eqnarray}
g_{ij}({\bm k}, {\bm R})= J_{ij}^{(2)}({\bm k}, {\bm R})- J_{ij}^{(1)}({\bm k}, {\bm R}),
\label{R224}
\end{eqnarray}
where
\begin{eqnarray}
J_{ij}^{(1)}({\bm k}, {\bm R}) &=&  i \int (k_p + K_{p}/2) \, G_\eta({\bm k} + {\bm  K} / 2) \, \overline{u_p ({\bm k} + {\bm  K} / 2 - {\bm  Q}) u_j(-{\bm k} + {\bm  K}  / 2)}
\nonumber\\
&&  \times  \meanB_i({\bm Q}) \exp(i {\bm K} {\bm \cdot} {\bm R})\,d {\bm  K} \,d {\bm  Q} ,
\label{QP1}\\
J_{ij}^{(2)}({\bm k}, {\bm R}) &=&  i \int (k_p + K_{p}/2) \, G_\eta({\bm k} + {\bm  K} / 2) \, \overline{u_i ({\bm k} + {\bm  K} / 2 - {\bm  Q}) u_j(-{\bm k} + {\bm  K}  / 2)}
\nonumber\\
&&  \times  \meanB_p({\bm Q}) \exp(i {\bm K} {\bm \cdot} {\bm R})\,d {\bm  K} \,d {\bm  Q} ,
\label{QQP1}
\end{eqnarray}
and the functions $J_{ij}^{(1)}$, $J_{ij}^{(2)}$, $G_\eta$ and $u_i$ depend also on $\omega$.
To simplify the notation, we do not show this dependence in the
following calculations.
First we determine the function $J_{ij}^{(1)}$, using new variables:
\begin{eqnarray}
\tilde {\bm k}_{1} &=& {\bm k} + {\bm  K} / 2 - {\bm  Q} \;, \quad
\tilde {\bm k}_{2} = - {\bm k} + {\bm  K} / 2 ,
\label{P24}\\
\tilde {\bm k} &=& (\tilde {\bm k}_{1} - \tilde {\bm k}_{2}) / 2 =
{\bm k} - {\bm  Q} / 2 \;, \; \tilde {\bm K} = \tilde {\bm k}_{1}
+ \tilde {\bm k}_{2} = {\bm  K} - {\bm  Q} .
\nonumber
\end{eqnarray}
Therefore,
\begin{eqnarray}
J_{ij}^{(1)}({\bm k}, {\bm R}) &=& i \int (k_{p} + K_{p}/2) \, f_{pj}({\bm k} - {\bm  Q} / 2, {\bm K} - {\bm  Q}) \, G_\eta({\bm k} + {\bm  K} / 2) \, \meanB_{i}({\bm  Q})
\nonumber\\
&&  \times \exp(i {\bm K} {\bm \cdot} {\bm R})\,d {\bm  K} \,d {\bm  Q} .
\label{QP2}
\end{eqnarray}
Since $|{\bm Q}| \ll |{\bm k}|$ and $|{\bm K}| \ll |{\bm k}|$, we use the Taylor expansion
\begin{eqnarray}
&& f_{pj}({\bm k} - {\bm Q}/2, {\bm  K} - {\bm  Q}) =
f_{pj}({\bm k},{\bm  K} - {\bm  Q}) - \frac{1}{2}
{\partial f_{pj}\over\partial k_s} Q_s  + O({\bm Q}^2) ,
\label{QP3}\\
&& G_\eta({\bm k} + {\bm  K} / 2) = G_\eta({\bm k}) \left[1 - \eta ({\bm k} \cdot {\bm  K}) G_\eta({\bm k}) \right] + O({\bm K}^2) .
\label{QQP3}
\end{eqnarray}
Substituting (\ref{QP3}) and~(\ref{QQP3}) into (\ref{QP2}), we get
\begin{eqnarray}
&& J_{ij}^{(1)}({\bm k}, {\bm R}) = i G_\eta({\bm k}) \int \biggl[\left(k_{p} - \eta G_\eta({\bm k}) k_{p}k_{s} K_{s} + {1 \over 2} K_{p} \right) \left(\int f_{pj}({\bm k},{\bm  K} - {\bm  Q}) \meanB_{i}({\bm Q}) \,d {\bm Q} \right)
\nonumber\\
&& \; - {1 \over 2} k_{p} \left(\int Q_s \meanB_{i}({\bm Q})
{\partial \over\partial k_s} f_{pj}({\bm k},{\bm  K} - {\bm  Q}) \,d {\bm Q} \right)\biggr]
 \exp(i {\bm K} {\bm \cdot} {\bm R}) \,d {\bm  K}
 + O({\bm K}^2, {\bm Q}^2, K_p Q_s) .
\nonumber\\
\label{QQP2}
\end{eqnarray}
We use the following identity:
\begin{eqnarray}
\nabla_{p} [f_{pj}({\bm k},{\bm R}) \meanB_{i}({\bm R})] = \int
i K_{p} \, [f_{pj}({\bm k},{\bm K}) \meanB_{i}({\bm K})]_{_{\bm K}}
\exp{(i {\bm K} {\bm \cdot} {\bm R})} \,d {\bm  K} ,
\label{QP4}
\end{eqnarray}
where
\begin{eqnarray}
[f_{pj}({\bm k},{\bm K}) \meanB_{i}({\bm K})]_{_{\bm K}} = \int
f_{pj}({\bm k},{\bm  K} - {\bm  Q}) \meanB_{i}({\bm Q}) \,d {\bm
Q} .
\label{QQP4}
\end{eqnarray}
Similarly,
\begin{eqnarray}
f_{pj}({\bm k},{\bm R}) \, \nabla_{p} \meanB_{i}({\bm R}) = \int \left[\int f_{pj}({\bm k},{\bm  K} - {\bm  Q}) \, i Q_{p} \meanB_{i}({\bm Q}) \, \,d {\bm Q} \right]_{_{\bm K}} \exp(i {\bm K} {\bm \cdot} {\bm R}) \,d {\bm  K} .
\label{QQP5}
\end{eqnarray}
Therefore, (\ref{QQP2})--(\ref{QQP5}) yield
\begin{eqnarray}
J_{ij}^{(1)}({\bm k},{\bm R}) &=& G_\eta \biggl\{ \left[i k_{p} \left(1+ i \eta \, G_\eta k_{s} \nabla_s \right) \left(\meanB_{i} \, f_{pj} \right)
+ \meanB_{i} \frac{1}{2} \nabla_p  f_{pj}\right]
\nonumber\\
&& + \frac{1}{2}\left(f_{pj} - k_{m} {\partial f_{mj}\over \partial k_p}\right) (\nabla_p \meanB_i) \biggr\} + O({\bm \nabla}^2)  .
\label{QP5}
\end{eqnarray}
A similar derivation is also performed for $J_{ij}^{(2)}$,
which yields
\begin{eqnarray}
J_{ij}^{(2)}({\bm k}, {\bm R}) &=& G_\eta \, \biggl\{\meanB_p \biggl[i k_p f_{ij}
+ {1 \over 2} \nabla_p f_{ij} - \eta k^2 G_\eta \, k_{sp} \nabla_s f_{ij}\biggr]
\nonumber\\
&& - (\nabla_s \meanB_p) \biggl[ {1 \over 2} k_p {\partial f_{ij} \over \partial k_s} + \eta k^2 G_\eta \, k_{sp} f_{ij}\biggr] \biggl\} + O({\bm \nabla}^2) .
\label{QQG2}
\end{eqnarray}
Therefore, (\ref{QP5}) and~(\ref{QQG2}) yield (\ref{G2}).

To determine the turbulent flux of particles for
small P\'{e}clet numbers, we rewrite (\ref{E1}) in Fourier space
using an equation similar to (\ref{R22}), and find
the solution of (\ref{E1}) as
\begin{eqnarray}
n({\bm k}, \omega) &=& - G_D({\bm k}, \omega) \,
i k_p \int N({\bm Q}) \, u_p({\bm k} - {\bm Q}, \omega) \,d{\bm Q},
 \label{EG1}
\end{eqnarray}
where $G_D({\bm k},\omega) = (D {\bm k}^2 + i \omega)^{-1}$.
The function $F_{j}({\bm k},{\bm R})$, defined by (\ref{EEE6}), is
\begin{eqnarray}
F_{j}({\bm k},{\bm R}) &=&  - i \int (k_p + K_{p}/2) \, \overline{u_p ({\bm k} + {\bm  K} / 2 - {\bm  Q}) u_j(-{\bm k} + {\bm  K}  / 2)}
\nonumber\\
&&  \times G_D({\bm k} + {\bm  K} / 2) \, N({\bm Q}) \exp(i {\bm K} {\bm \cdot} {\bm R})\,d {\bm  K} \,d {\bm  Q} ,
\label{AAQP1}
\end{eqnarray}
where the functions $F_{j}$, $G_D$ and $u_i$ depend also on $\omega$, and
$N$ depend on $t$ as well.
To simplify the notation, we do not show this dependence here.
We perform calculations similar to those in (\ref{QP1})--(\ref{QP5}).
In particular, after the Taylor expansion for $|{\bm Q}| \ll |{\bm k}|$
and $|{\bm K}| \ll |{\bm k}|$, we arrive at expression (\ref{E5}) for the
turbulent flux of particles in Fourier space for small P\'{e}clet numbers.

\section{Derivation of (\ref{C4}) for Rm $\ll 1$}
\label{appC}

The effective pumping velocity $\bm{\gamma}$ of the large-scale magnetic field
is independent of ${\bm \lambda}$ for Rm $\ll 1$.
Indeed, the contribution of ${\bm \lambda}$ to the effective pumping velocity $\bm{\gamma}$ can only arise from the term
\begin{eqnarray}
g_{ij}^{(1)}({\bm k},\omega) = i k_p G_\eta(k,\omega)
[\meanB_p f_{ij}({\bm k},\omega) - \meanB_i f_{pj}({\bm k},\omega)].
\label{G502}
\end{eqnarray}
Integration over solid angles in ${\bm k}$ space in $g_{ij}^{(1)}=\int g_{ij}^{(1)}({\bm k},\omega) \,d{\bm k} \,d \omega$ yields
\begin{eqnarray}
g_{ij}^{(1)} = - {\int \,dk  E(k) \int  \,d \omega \Phi_u(\omega) G_\eta(k,\omega)  \over 6 \,(1+ \sigma_c)} \left[\meanB_i \lambda_j + \meanB_j \lambda_i - \left(\sigma_c+{1\over 2} \right) (\meanB_i \nabla_j + \meanB_j \nabla_i)\overline{{\bm u}^2}\right],
\nonumber\\
\label{G500}
\end{eqnarray}
which is a symmetric tensor with respect to the indexes $i$ and $j$.
Therefore, it cannot contribute to the mean electromotive force
nor to $\bm{\gamma}$.
For integration over solid angles in ${\bm k}$ space in the expression for
$g_{ij}^{(1)}$, we take into account that $\int k_{ij} \sin \theta \,d\theta
\,d\varphi =(4 \pi/3) \delta_{ij}$, where we use spherical coordinates
$(k, \theta, \varphi)$ in ${\bm k}$ space.

The remaining contribution, which is proportional to the mean magnetic field, is
\begin{eqnarray}
&& g_{ij}^{(2)}({\bm k},\omega) =
G_\eta \, \biggl[\meanB_p \biggl(
{1 \over 2} \nabla_p f_{ij} - \eta k^2 G_\eta \, k_{sp} \nabla_s f_{ij}\biggr)
- \meanB_i \biggl({1 \over 2} \nabla_s  - \eta k^2 G_\eta \, k_{sp} \nabla_p  \biggr) f_{sj} \biggr].
\nonumber\\
\label{G501}
\end{eqnarray}
Since we neglect effects $\sim$O$(
\lambda_i \nabla_i\overline{{\bm u}^2}, \lambda^2 \, \overline{{\bm u}^2}, \nabla^2 \, \overline{{\bm u}^2})$,
the term $g_{ij}^{(2)}({\bm k},\omega)$ describes only
the contribution to $\bm{\gamma}$ caused by ${\bm \nabla}\overline{{\bm u}^2}$,
i.e., the effect of inhomogeneity of turbulence, rather than the effect of stratification on $\bm{\gamma}$.
A non-zero contribution to $g_{ij}^{(2)}$ (and to the mean electromotive force) is only from
the last two terms in (\ref{G501}), which are proportional to $f_{sj}$, where
$f_{sj} \propto (\delta_{sj} - k_{sj} + 2 \sigma_c \, k_{sj}) \overline{{\bm u}^2}$.
Indeed, the first two terms $\propto f_{ij}$ in (\ref{G501})
(which is a symmetric tensor with respect to the indexes $i$ and $j$)
cannot contribute to the electromotive force.
Here we also took into account that $\int k_{i} \sin \theta \,d\theta \,d\varphi =0$.
After integration in ${\bm k}$ and $\omega$ space we arrive at (\ref{C4})
for the effective pumping velocity $\bm{\gamma}$ for Rm $\ll 1$.
For integration over $\omega$, we use the following integrals $\tau_0 \gg (\eta k^2)^{-1}$:
\begin{eqnarray}
&& \int_{-\infty}^{\infty} {d\omega \over (\pm i \omega + \eta k^2)
\, (\omega^2 + \tau_0^{-2})} = {\pi \, \tau_0 \over \tau_0^{-1} + \eta \,
k^2}  \approx {\pi \, \tau_0 \over \eta \, k^2} ,
\label{J28}\\
&& \int_{-\infty}^{\infty} {d\omega \over (i \omega + \eta k^2)
\, (- i \omega + \eta k^2) \, (\omega^2 +
\tau_0^{-2})} ={\pi \, \tau_0 \over \eta \, k^2 \left(\tau_0^{-1} + \eta \, k^2\right)} \approx {\pi \, \tau_0 \over \eta^2 \,
k^4} .
\label{J29}
\end{eqnarray}

Applying the turbulence model of \cite{amir2017} for arbitrary stratification
\citep[see also][]{elperin1995},
\begin{eqnarray}
\overline{u_i({\bm k}) \, u_j(-{\bm k})} = {\overline{{\bm u}^2} \, E(k) \over 8 \pi (k^2 + \lambda^2)} \left[\delta_{ij} - {k_i \, k_j \over k^2} + {i \over k^2} \, \big(\lambda_i \, k_j - \lambda_j \, k_i\big) + {\lambda^2 \over k^2} \left(\delta_{ij} - {\lambda_i \, \lambda_j \over \lambda^2}\right)\right] ,
\nonumber\\
\label{G600}
\end{eqnarray}
we show that there are no contributions of $\bm{\lambda}$ to the effective
pumping velocity $\bm{\gamma}$ of the large-scale magnetic field---even for strong stratifications.
In particular, we substitute (\ref{G600}) into (\ref{G500}), taking into account that
$\int k_{p} \sin \theta \,d\theta \,d\varphi =0$.
Note also that a cross-effect of stratification and
inhomogeneity of turbulence, $(\lambda_i \nabla_j  - \lambda_j \nabla_i) \overline{\uu^2}$,
in the model of the background turbulence does not contribute
to the effective pumping velocity $\boldsymbol\gamma$ for the same reasons.

\section{Derivation of (\ref{D8}) for Rm $\gg 1$ and (\ref{T1}) for Pe $\gg 1$}
\label{appB}

For the derivation of (\ref{D8}), we perform
the following calculation.
We use the identity:
\begin{eqnarray}
{\partial \over \partial t} \overline{b_i({\bm k}_1,t) \,
u_j({\bm k}_2,t)} = \overline{{\partial b_i({\bm k}_1,t) \over \partial t} \,
u_j({\bm k}_2,t)} + \overline{b_i({\bm k}_1,t) \, {\partial u_j({\bm k}_2,t) \over \partial t}
} .
\label{AL1}
\end{eqnarray}
We rewrite the induction equation~(\ref{D2}) for magnetic fluctuations in ${\bm k}$ space:
\begin{eqnarray}
{\partial b_i({\bm k}) \over \partial t} &=& i k_p \, \left(\int \overline{B}_p({\bm Q}) \,
u_i({\bm k}-{\bm Q}) \,d{\bm Q} - \int \overline{B}_i({\bm Q}) \,  u_p({\bm k}-{\bm Q}) \,d{\bm Q} \right) + b_i^{\rm(N)}({\bm k}) ,
\label{AL2}
\end{eqnarray}
using (\ref{R22})--(\ref{R223}).
Here $b_i^{\rm(N)}({\bm k})$ includes the nonlinear terms.
For brevity of notations we omit below the variable $t$ in the functions
$\overline{B}_i({\bm Q},t)$, $b_i({\bm k},t)$, $b_i^{\rm(N)}({\bm k},t)$ and $u_i({\bm k},t)$.
To derive (\ref{D8}), we use (\ref{AL1}) and (\ref{AL2}) which yield
\begin{eqnarray}
{\partial g_{ij}({\bm k},{\bm R}) \over \partial t} &=& \tilde  J_{ij}^{(1)}({\bm k},{\bm R}) -\tilde  J_{ij}^{(2)}({\bm k},{\bm R}) + g_{ij}^{\rm(N)}({\bm k},{\bm R}) ,
\label{ALLL2}
\end{eqnarray}
where
\begin{eqnarray}
\tilde  J_{ij}^{(1)}({\bm k},{\bm R}) &=&  i \int (k_p + K_{p}/2)
\, \meanB_p({\bm Q}) \, \overline{u_i ({\bm k} + {\bm  K} / 2 - {\bm  Q})
u_j(-{\bm k} + {\bm  K}  / 2)}
\nonumber\\
&&  \times \exp(i {\bm K} {\bm \cdot} {\bm R})\,d {\bm  K} \,d {\bm  Q} ,
\label{MP1}\\
\tilde  J_{ij}^{(2)}({\bm k},{\bm R}) &=&  i \int (k_p + K_{p}/2)
\, \meanB_i({\bm Q}) \, \overline{u_p ({\bm k} + {\bm  K} / 2 - {\bm  Q})
u_j(-{\bm k} + {\bm  K}  / 2)}
\nonumber\\
&&  \times \exp(i {\bm K} {\bm \cdot} {\bm R})\,d {\bm  K} \,d {\bm  Q} .
\label{MP2}
\end{eqnarray}
Next, we perform calculations that are similar to (\ref{P24})--(\ref{QP3})
and (\ref{QQP2})--(\ref{QP5}).
In particular, we introduce new variables:
$\tilde {\bm k} = (\tilde {\bm k}_{1} - \tilde {\bm k}_{2}) / 2=
{\bm k} - {\bm  Q}/2$ and $\tilde {\bm K} = \tilde {\bm k}_{1} + \tilde {\bm k}_{2}
= {\bm  K} - {\bm  Q}$, and use the Taylor expansion for $|{\bm Q}| \ll |{\bm k}|$ and $|{\bm K}| \ll |{\bm k}|$, which yield
\begin{eqnarray}
\tilde J_{ij}^{(1)}({\bm k},{\bm R}) &=& \meanB_p \left(i k_{p} f_{ij}+ \frac{1}{2} \nabla_p  f_{ij}\right) - \frac{1}{2} k_{m} {\partial f_{ij}\over
\partial k_p} \, \nabla_p \meanB_m + O({\bm \nabla}^2),
\label{MP4}\\
\tilde J_{ij}^{(2)}({\bm k},{\bm R}) &=& \meanB_i \left(i k_{p} f_{pj}+ \frac{1}{2} \nabla_p  f_{pj}\right) + \frac{1}{2}\left(f_{pj} - k_{m} {\partial f_{mj}\over
\partial k_p}\right) (\nabla_p \meanB_i) + O({\bm \nabla}^2).
\label{MP5}
\end{eqnarray}
Therefore, (\ref{AL2})--(\ref{MP5}) yield
(\ref{D8}) for Rm $\gg 1$.

For the derivation of (\ref{T1}), we perform
the following calculation.
We use the identity:
\begin{eqnarray}
{\partial \over \partial t} \overline{n({\bm k}_1,t) \,
u_j({\bm k}_2,t)} = \overline{{\partial n({\bm k}_1,t) \over \partial t} \,
u_j({\bm k}_2,t)} + \overline{n({\bm k}_1,t) \, {\partial u_j({\bm k}_2,t) \over \partial t}} .
\label{AAL1}
\end{eqnarray}
We rewrite equation~(\ref{E1}) for fluctuations of the particle number density, in ${\bm k}$ space:
\begin{eqnarray}
{\partial n({\bm k}) \over \partial t} &=& - i k_p \, \int N({\bm Q}) \,
u_p({\bm k}-{\bm Q}) \,d{\bm Q}  + n^{\rm(N)}({\bm k}) ,
\label{AAL2}
\end{eqnarray}
using (\ref{R22}). Here $n^{\rm(N)}({\bm k})$ includes the nonlinear terms. For brevity of notations we omit below the variable $t$ in the functions
$N({\bm Q},t)$, $n({\bm k},t)$, $n^{\rm(N)}({\bm k},t)$ and $u_i({\bm k},t)$.
To derive (\ref{T1}), we use (\ref{AAL1}) and (\ref{AAL2}) which yield
\begin{eqnarray}
{\partial F_{j}({\bm k},{\bm R}) \over \partial t} &=& \tilde  J_{j}^{(3)}({\bm k},{\bm R}) + F_{j}^{\rm(N)}({\bm k},{\bm R}) ,
\label{ALLL2}
\end{eqnarray}
where
\begin{eqnarray}
\tilde  J_{j}^{(3)}({\bm k},{\bm R}) &=&  - i \int ({\bm k}_p + K_{p}/2)
\, N({\bm Q}) \, \overline{u_p ({\bm k} + {\bm  K} / 2 - {\bm  Q})
u_j(-{\bm k} + {\bm  K}  / 2)}
\nonumber\\
&&  \times \exp(i {\bm K} {\bm \cdot} {\bm R})\,d {\bm  K} \,d {\bm  Q} .
\label{AMP1}
\end{eqnarray}
We perform calculations that are similar to (\ref{P24})--(\ref{QP3})
and (\ref{QQP2})--(\ref{QP5}), which yield
\begin{eqnarray}
\tilde  J_{j}^{(3)}({\bm k},{\bm R}) = - N \left(i k_p f_{pj}+ {1 \over 2} \nabla_p f_{pj} \right) - {1 \over 2} \left(f_{pj} - k_{m} \, {\partial f_{mj} \over \partial k_{p}}
\right) \nabla_p N .
\label{AT1}
\end{eqnarray}
Therefore, (\ref{AAL1})--(\ref{AT1}) yield (\ref{T1}) for Pe $\gg 1$.

\bibliographystyle{jpp}

\bibliography{paper-JPP}

\end{document}